\documentclass{ws-ijmpa}
\usepackage[super, compress]{cite}
\usepackage{physics}
\usepackage{multirow}
\usepackage{multicol}
\usepackage{booktabs}
\usepackage{xcolor}
\usepackage{longtable}
\usepackage[verbose,hypertexnames=false]{hyperref}
\hypersetup{colorlinks=false,allbordercolors=blue,pdfborderstyle={/S/U/W 1}}
\begin{document}
	
	\markboth{}{}
%	
%	 Publisher's Area please ignore 
%	
%	\catchline{}{}{}{}{}
%	
%	
	
	\title{Sensitivity of Two-Body Non-Leptonic Branching Fractions to Theoretical Mass Variations in Heavy-Light Mesons}

	\author{Manakkumar Parmar and Ajay Kumar Rai}
	
	\address{Department of Physics, Sardar Vallabhbhai National Institute of Technology,\\
		Surat, Gujarat 395 007, India\\
		mgp.physica@gmail.com \textup{and} raiajayk@gmail.com}

	\maketitle
	
	% \begin{history}
	% 	\received{Day Month Year}
	% 	\revised{Day Month Year}
	% 	\accepted{Day Month Year}
	% 	\published{Day Month Year}
	% \end{history}
	
	\begin{abstract}
		This study investigates the sensitivity of two-body non-leptonic branching fractions to theoretical mass variations in heavy-light mesons ($D$, $D_s$, $B$, and $B_s$). Utilizing the factorization framework, we compare predictions derived from phenomenological masses evaluated with Gaussian and hydrogenic wavefunctions. For bottom meson decays, naive factorization with the number of color $N = 3$ aligns well with experimental data, and the $N \to \infty$ limit offers no improvement. Furthermore, the theoretical mass variation between wavefunction models induces a pronounced, non-linear sensitivity in the branching fractions, establishing the accurate Gaussian mass as a crucial baseline. Conversely, in the charm sector, naive factorization is inherently limited by final-state interactions due to insufficient relativistic recoil. While the $N \to \infty$ limit partially compensates for this, the systematically lower hydrogenic mass yields more accurate rates for several color-suppressed channels. This mass underestimation acts as a necessary kinematic regulator, cleanly offsetting the inflated amplitudes inherent to charm factorization. Ultimately, combining reliable Gaussian mass predictions with factorization provides a simple formalism extendable to the decay properties of unobserved exotics, such as excited $B_c$ mesons and $T_{bb}$ tetraquarks.
	\end{abstract}
	
	
	\ccode{PACS numbers: TBD}
	
	\section{Introduction}
         In theory, nonleptonic decays are more complicated than leptonic or semileptonic ones because they involve local four-quark interactions. Usually, these decays are handled using the factorization approach, where the decay amplitude is written as a product of the meson decay constant and weak transition form factors. The idea behind this method comes from Bjorken’s color transparency concept\cite{BJORKEN1989325}, which suggests that in energetic B decays, the outgoing light meson moves quickly in the opposite direction to the other meson, almost avoiding the color field of the parent particle. This makes the factorization approximation valid.
         
         In this paper, we use the form factors derived through quasi-potential approach. The numerical values of the form factors are available in Refs. \citen{PhysRevD.101.013004, PhysRevD.106.013004}. Form factors predicted are valid across the kinematically allowed region. Their validity has been thoroughly tested against experimental data through various calculated semi-leptonic decay observables, such as branching fractions, forward-backward asymmetries, and polarizations. Related studies on applying these form factors to nonleptonic decays can be found in Refs. \citen{PhysRevD.56.312, PhysRevD.87.094028, PhysRevD.98.074512}. Additionally, a recent investigation of charmless two-body $B$ meson decays employs the perturbative QCD factorization approach in ref.\citen{Chai:2022ptk}. Earlier works are available in Refs. \citen{PhysRevD.85.094003, PhysRevD.103.056006, PhysRevD.76.074018}.

        It's commonly accepted that combining Coulomb and linear potentials within a non-relativistic framework accurately describes the spectrum of heavy mesons. That said, there's been ongoing uncertainty about whether this works equally well for mesons made of heavy-light quark pairs. The objective of this paper is to explore how different theoretical mass inputs alter the predictions of the naive factorization approach for heavy-light meson decays.

        This paper is organized as follows. Section 2 establishes the theoretical framework by introducing the hydrogenic and Gaussian wavefunctions and detailing the Hamiltonian for the $D$, $D_s$, $B$, and $B_s$ mesons. In Section 3, the role of the Wilson coefficients is discussed, alongside the general expressions for the transition matrix elements and the necessary input parameters. Section 4 provides the explicit decay amplitude formulas employed in this study. The numerical results are presented in Section 5, where the predicted branching fractions are compared against the latest experimental data from the Particle Data Group (PDG 2024) and prior theoretical literature. Finally, concluding remarks are provided in Section 6.
        
	\section{Theoretical Framework}	
    	In this section, we establish the theoretical framework used to investigate the nonleptonic two-body decays of heavy-light mesons under the factorization approximation. This method simplifies the calculation by expressing the decay amplitude as a product of more manageable hadronic matrix elements. The factorization assumption is supported by the concept of color transparency, where the rapidly moving light meson quickly exits the parent meson's color field. This enables the weak decay amplitude to be approximated as a combination of a meson decay constant and a weak transition form factor. For mesons composed of a heavy and a light quark—such as the $D$, $D_s$, $B$ and $B_s$ a relativistic Hamiltonian is used to account for the relativistic motion of the constituent quarks.\cite{devlani_spectroscopy_2011, PhysRevD.51.168, HWANG1997117}
	
	\begin{equation}
		\mathcal{H} = \sqrt{\vec{p}^2 + m_Q^2} + \sqrt{\vec{p}^2 + m_{\bar{q}}^2} + U(r),
		\label{hamiltonian}
	\end{equation}
	where $\vec{p}$ is the relative momentum of the quark-antiquark and $m_Q$ and $m_{\bar{q}}$ are the heavy-quark and light-quark masses respectively. The Hamiltonian in Eq.(\ref{hamiltonian}) is the energy of the meson in the meson rest frame. $U(r)$ is the quark-antiquark potential\cite{PhysRevC.78.055202, Rai_2005, AjayKRai_2002},
	
	\begin{equation}
		U(r) = -\frac{\alpha_c}{r} + Ar^\nu + U_0,
	\end{equation}
	where $A$ is the potential parameter and $\nu$ is a general power index, such that the choice of $\nu = 1$ corresponds to the Coulomb plus linear potential with a constant term $U_0$. $\alpha_c = \frac{4}{3} \alpha_s,$ where $\alpha_s$ represents the strong running coupling constant. The QCD coupling $\alpha_s$ is evaluated using a basic model that incorporates coupling freezing\cite{PhysRevD.70.016007, simonov1993perturbationtheorynonperturbativeqcd}.
	
	\begin{equation}
		\alpha_s(M^2) = \frac{4\pi}{(11 - \frac{2}{3}n_f) \ln \frac{M^2 + M_B^2}{\Lambda^2}},
	\end{equation}
	where the scale is taken as $M = 2m_Qm_{\bar{q}}/(m_Q + m_{\bar{q}})$, the background mass is $M_B = 2.24 \sqrt{A'} = 0.95$ GeV, $\Lambda = 413$ MeV\cite{PhysRevD.70.016007, simonov1993perturbationtheorynonperturbativeqcd, PhysRevD.79.114029}.\\	
	In this study, both Gaussian and hydrogen-like wave functions have been employed. The Gaussian wave function in position space is given by\cite{devlani_mass_2013}	
	
	\begin{equation}
		R_{nl}(\gamma, r) = \gamma^{3/2} \sqrt{\frac{2(n-1)!}{\Gamma(n+l+1/2)}} (\gamma r)^l e^{-\gamma^2 r^2 / 2} L_{n-1}^{l+1/2}(\gamma^2 r^2),
	\end{equation}
	and the hydrogen-like wave function has the form\cite{devlani_spectroscopy_2012}
	
	\begin{equation}
		R_{nl}(\gamma, r) = \left( \frac{\gamma^3 (n-l-1)!}{2n(n+l)!} \right)^{1/2} (\gamma r)^l e^{-\gamma r / 2} L_{n-l-1}^{2l+1}(\gamma r ).
	\end{equation}
	Here, $\gamma$ is the variational parameter and $L$ is the Laguerre polynomial. For this analysis, the phenomenological potential index is fixed at $\nu = 1.0$. The state-dependent variational parameters ($\gamma$) are determined via the Virial theorem. For the ground ($1S$) states, the optimized values of $\gamma$ for the hydrogenic and Gaussian wave functions, respectively, are: $1.150$ GeV and $0.448$ GeV for the $D$ meson; $1.203$ GeV and $0.467$ GeV for the $D_s$ meson; $1.247$ GeV and $0.488$ GeV for the $B$ meson ; and $1.320$ GeV and $0.515$ GeV for the $B_s$ meson. For more details one can refer to Refs.\citen{devlani_mass_2013,devlani_spectroscopy_2011, devlani_spectroscopy_2012}.
		\subsection{Effective Hamiltonians and Factorization}
				Weak decays of heavy flavored hadrons, such as those involving $b$ and $c$ quarks, provide critical probes of beyond-Standard-Model physics through anomalies in lepton universality and rare decay channels observed at LHCb and Belle experiments\cite{dicanto2022weakdecaysbc, Sun:2015faa}.
				
				The Cabibbo–Kobayashi–Maskawa (CKM) matrix plays a central role in the Standard Model of particle physics by describing the mixing of quark flavors under weak interactions. Specifically, it encodes the mismatch between the mass eigenstates and weak interaction eigenstates of the three generations of quarks. When a down-type quark undergoes a weak charged current interaction, such as during a beta decay process, it transits into an up-type quark with a probability amplitude given by an element of the CKM matrix. This $3 \times 3$ unitary matrix introduces complex phases and mixing angles that not only quantify the strength of flavor-changing transitions but also serve as the only source of CP violation in the quark sector within the Standard Model. 
			
				The unitarity of the CKM matrix imposes stringent constraints on the relationships among its elements, often visualized through unitarity triangles. These triangles offer valuable insight into CP-violating phenomena and are critical to testing the consistency of the Standard Model\cite{PDG2024}. Precision measurements of the CKM elements, both directly from decay rates and indirectly from global fits, continue to serve as a powerful probe for potential physics beyond the Standard Model.
			
				For both charm and bottom meson decays, the starting point is the effective weak Hamiltonian expressed in terms of local four-quark operators. In the case of charm decays, the Hamiltonian can be written as\cite{Galkin2023}	
			\begin{align}
				\mathcal{H}_{\text{eff}} &=\frac{G_F}{\sqrt{2}} V^*_{cq_1}V_{uq_2} \big[c_1(\mu) (\bar{q}_{1\alpha} c_\alpha)_{V-A} (\bar{u}_\beta q_{2\beta})_{V-A} \nonumber \\
				& \quad+ c_2(\mu) (\bar{q}_{1\alpha} c_\beta)_{V-A} (\bar{u}_\beta q_{2\alpha})_{V-A} 
				\big]
			\end{align}
			For the processes $\bar{b} \to \bar{c} u \bar{d}$, $\bar{b} \to \bar{c} u \bar{s}$, $\bar{b} \to \bar{u} c \bar{d}$, and $\bar{b} \to \bar{u} c \bar{s}$, the effective Hamiltonian reads\cite{Galkin2023}
			\begin{equation}
				\mathcal{H}_{\text{eff}} = \frac{G_F}{\sqrt{2}} [V_{qb}^* V_{q_1 q_2} (c_1(\mu) O_1 + c_2(\mu) O_2)].
			\end{equation}
			where \(G_F\) is the Fermi constant which is equal to $1.166\times 10^{-5}$ GeV$^{-2}$, where $V_{q_1 q_2}$ represent the corresponding CKM matrix elements. We use the CKM matrix in the Wolfenstein parameters with central values, $\lambda=0.22501$, $A=0.826$, $\bar{\rho}=0.1591$, and $\bar{\eta}=0.3523$\cite{PDG2024} and the \(c_i(\mu)\) are Wilson coefficients.\\
			For transitions, $\bar{b} \to \bar{c} c \bar{d}$, $\bar{b} \to \bar{c} c \bar{s}$, $\bar{b} \to \bar{u} u \bar{d}$ and $\bar{b} \to \bar{u} u \bar{s}$, the complete effective Hamiltonian consists of the previously stated tree-level operators plus the penguin contributions, which are given by \cite{Galkin2023}:
			\begin{equation}
				\mathcal{H}_{\text{penguin}} = -\frac{G_F}{\sqrt{2}} V_{tb}^* V_{tq} \sum_{i=3}^{10} c_i(\mu) O_i.
			\end{equation}
			where $O_i$ are the local four-quark operators (detailed explicit expressions can be found in Ref. \citen{Galkin2023}), and $c_i(\mu)$ are the corresponding Wilson coefficients evaluated at the renormalization scale $\mu \approx m_b$. All other symbols retain their standard definitions.
			
	\section{$c_i$'s Wilson Coefficients}
			Wilson coefficients \( c_i(\mu) \) encode the short-distance QCD dynamics and allow the separation of high- and low-energy scales via the operator product expansion. The effective weak Hamiltonian is constructed by evolving these coefficients from the electroweak scale down to a characteristic hadronic scale, taken to be \( \mu \approx m_c \) for charm decays and \( \mu \approx m_b \) for bottom decays.
			\begin{table}[h!]
				\tbl{Wilson coefficients at scales $\mu \approx m_c$\cite{PhysRevD.53.2506} (charmed mesons) and $\mu \approx m_b$\cite{RevModPhys.68.1125} (bottom mesons). \label{wilson-table}}
				{\begin{tabular}{@{}lccccccccccc@{}}
						\toprule
						\textbf{Scale} & $c_1$ & $c_2$ & $c_3$ & $c_4$ & $c_5$ & $c_6$ & $c_7/\alpha$ & $c_8/\alpha$ & $c_9/\alpha$ & $c_{10}/\alpha$ \\
						\midrule
						$\mu \approx m_c$ & 1.26 & $-0.51$ & -- & -- & -- & -- & -- & -- & -- & -- \\
						$\mu \approx m_b$ & 1.105 & $-0.228$ & 0.013 & $-0.029$ & 0.009 & $-0.033$ & 0.005 & 0.060 & $-1.283$ & 0.266 \\
						\bottomrule
				\end{tabular}} 
			\end{table}
			For charm meson decays, only the current–current operators are retained, as penguin contributions are heavily suppressed.\\
			For bottom decays, the analysis includes both tree-level and penguin operators, reflecting the more prominent role of loop-induced processes in \(B\)-physics. The full set of Wilson coefficients \( c_1 \) through \( c_{10} \) is employed, with electroweak penguin coefficients normalized by the electromagnetic coupling \( \alpha \). At \( \mu = m_b \), the values are as shown in the Table~\ref{wilson-table}. These coefficients are then used to construct effective coefficients \( a_i \), defined to include color-suppressed terms:
			\begin{equation}
				a_i = \begin{cases}
					c_i + c_{i+1}/N, & \text{for odd } i, \\
					c_i + c_{i-1}/N, & \text{for even } i.
				\end{cases}
			\end{equation}
			The numerical values and scale dependence of these coefficients play a central role in shaping the predicted decay amplitudes. Moreover, their inclusion is essential in quantifying the relative contributions of tree-level and penguin topologies, especially in processes where interference effects are non-negligible. The careful treatment of these coefficients, along with the large-\( N \) limit used to approximate nonfactorizable corrections.\\
			For the nonleptonic weak decay, when we use the factorization approach, there is a identity known as Fierz transformation that allows us to rearrange the structure of the operators, particularly in terms of how the color and spinor indices are contracted. Thus, we can separate the decay amplitude into simpler products of matrix elements. 
			The transformation is:
			\begin{equation}
				\delta_{lj}\delta_{ik}=\frac{1}{N}\delta_{ij}\delta_{lk}+2T^a_{lk}T^a_{ij},
			\end{equation}
			where, $i$, $j$ $k$ and $l$ are color indices, $T^a$ are the SU(3) color matrices (Gell-Mann matries), and $N$ is the number of colors in QCD (typically $N=3$). By applying this transformation to four-quark terms like:
			\begin{equation}
				(s\bar{s})_{V-A}(\bar{u}c)_{V-A},
			\end{equation}
			the expression is rewritten in a way that separates a color-singlet part (proportional to $\frac{1}{N}$) and a color-octet part (nonfactorizable, involving the $T^a$matrices). In this approximation, part of the effects of nonfactorizable interactions are empirically absorbed by adjusting the color factor $N\to\infty$ which improves the accuracy with the experimental data.\\
            In the factorization approximation, the nonfactorizable terms—associated with soft gluon exchanges and final-state interactions—are neglected, though they are sometimes partly compensated empirically by taking the limit $N \to \infty$ \cite{Galkin2023}, where $N$ is the number of colors. However, this large-$N$ limit is not universally applicable; while it improves theoretical predictions in the charm sector, naive factorization with $N=3$ remains more accurate for bottom meson decays.  
			
            These loop-induced diagrams introduce further operators that, after Fierz rearrangements, yield terms proportional to the Wilson coefficients of both even and odd parity operators. The interplay between these different contributions is essential to correctly predict branching fractions, especially since in some bottom decays, the penguin terms can be comparable to or even exceed the tree-level ones.
	
		\subsection{Hadronic Matrix Elements and Input Parameters}
			Within the factorization framework, the hadronic matrix elements are factorized into two parts. One is the decay constant, defined as the matrix element of the weak current between the vacuum and the produced meson. For example, for a pseudoscalar meson \(P\), the definition is,
			\begin{equation}
				\langle P(p) | (\bar{q}_1 q_2)_{V-A} | 0 \rangle = -i f_P p^\mu
			\end{equation}
			\begin{equation}
				\bra{V}(\bar{q}_1 q_2)_{V-A}\ket{0} =  i f_V m_V \epsilon_\mu^*
			\end{equation}
			and similarly, for a vector meson \(V\), the decay constant \(f_V\) is defined via the vector current. The other essential ingredient is the transition form factor, which parameterizes the matrix element of the weak current between the initial and final meson states. These form factors, denoted \(f_0(q^2)\), \(f_+(q^2)\) for pseudoscalar transitions or \(V(q^2), A_0(q^2), A_1(q^2), A_2(q^2)\) for vector transitions, are obtained from a quark model based on the quasi-potential approach.\\
			For the process where meson $M$ decays into two pseudoscalar mesons, i.e., $M\to P_1,P_2$, one has\cite{Galkin2023},
			\begin{equation}
				\mathcal{M}_{M\to P_1,P_2} = - i f_{P_2}(m_M^2-m_{P_1}^2)f_0(m_{P_2}^2)
				\label{PP}
			\end{equation}
			for $M \to P, V$,
			\begin{equation}
				\mathcal{M}_{M\to P, V} = 2if_V f_+(m_V^2)m_M\abs{\vec{p}}
				\label{PV}
			\end{equation}
			for $M \to V, P$,
			\begin{equation}
				\mathcal{M}_{M\to V, P} = 2if_P A_0(m_P^2)m_M\abs{\vec{p}}
			\end{equation}
			for $M\to V_1, V_2$,
			\begin{equation}
				\abs{\mathcal{M}_{M\to V_1,V_2}}^2 = f_{V_2}^2 m_{V_2}^2(\abs{H_0}^2+\abs{H_+}^2+\abs{H_-}^2),
			\end{equation}
			where 
			\begin{equation}
				H_\pm = -(m_M+m_{V_1})A_1(m_{V_2}^2) \pm \frac{2m_M \abs{\vec{p}}}{m_M+m_{V_1}}V(m_{V_2}^2),
			\end{equation}
			and
			\begin{align}
				H_0 &= -(m_M+m_{V_1}) A_1(m_{V_2}^2)\frac{m_M^2-m_{V_1}^2-m_{V_2}^2}{2 m_{V_1} m_{V_2}} \nonumber\\
				&\quad +\frac{2m_M^2\abs{\vec{p}}^2}{(m_M+m_{V_1})m_{V_1}m_{V_2}}A_2(m_{V_2}^2).
			\end{align}
			The momentum can be calculated using the two-body decay momentum formula:
			\begin{equation}
				\abs{\vec{p}} = \lambda^{1/2}(m_M^2, m_1^2, m_2^2)/(2m_M),
				\label{kallen}
			\end{equation}
			where $\lambda(x,y,z)=x^2+y^2+z^2-2(xy+yz+zx)$ is a Källén function.\\
			The values of the decay constants used (in MeV) are provided in Table 2. For more details, please refer to Refs. \cite{rosner_leptonic_2016, EBERT200693, PhysRevD.98.074512, PhysRevC.60.055214, PhysRevD.71.014029, PhysRevD.75.054004, PhysRevD.60.111502}\\
			\begin{table}[h!]
				\tbl{Decay constants \(f\) for various mesons (in MeV)\cite{Galkin2023}} 
				{\begin{tabular}{ccccccccccc}
					\toprule
					\( f_\pi \) & \( f_K \) & \( f_{K^*} \) & \( f_\eta^u \) & \( f_\eta^s \) & \( f^u_{\eta'} \) & \( f^s_{\eta'} \) & \( f_\rho \) & \( f_\omega \) & \( f_\phi \) \\
					\midrule
					130 & 155 & 217 & 78 & -112 & 63 & 137 & 205 & 187 & 215 \\
					\botrule
				\end{tabular}}			
			\end{table}
			\begin{table}[h]
				\tbl{Masses used in our calculation and their comparison with experimental(in GeV)}
				{\begin{tabular}{lccc}
						\toprule
						\multicolumn{1}{c}{\multirow{2}[4]{*}{Particle}} & \multicolumn{2}{c}{Mass \cite{devlani_mass_2013, devlani_spectroscopy_2011, devlani_spectroscopy_2012}} & \multirow{2}[4]{*}{Experimental\cite{PDG2024}} \\
						\cmidrule{2-3}      & Hydrogenic & Gaussian &  \\
						\midrule
						$ D^{\pm} $ & 1.702 & 1.865 & 1.869 \\
						$ D^{0} $ & 1.702 & 1.865 & 1.864 \\
						$ D_s^{\pm} $ & 1.801 & 1.970  & 1.968 \\
						$ B^{\pm, 0} $ & 5.146 & 5.266 & 5.279 \\
						$ B_s^0 $ & 5.236 & 5.355 & 5.366 \\
						\bottomrule
					\end{tabular}}
			\end{table}
			Finally, the formula for the branching fraction is:
			\begin{equation}
				\mathcal{B}=\tau\frac{\abs{\vec{p}}}{8\pi m^2}\abs{\mathcal{A}}^2,
				\label{Bfeqn}
			\end{equation}
			where $\tau$ and $m$ are lifetime and mass of the parent particle, respectively. $|\vec{p}|$ is 3-momentum and $\mathcal{A}$ is amplitude.
			In this work, we adopt the form factors for meson weak transitions as constructed in Refs. \citen{Galkin2023, PhysRevD.101.013004, PhysRevD.106.013004}. These form factors provide a consistent and physically motivated description of the nonperturbative QCD dynamics involved in hadronic transitions. The approach systematically accounts for the correct behavior of meson wavefunctions when viewed from different frames and includes effects from intermediate energy states. The resulting form factors are expressed through formulas valid across the entire kinematically allowed range of momentum transfer with formulas:
			\begin{itemlist}
				\item for $f_+(q^2)$, $V(q^2)$ and $A_0(q^2)$\cite{PhysRevD.101.013004}:
				\begin{equation}
					F(q^2)=\frac{F(0)}{\pqty{1-\frac{q^2}{M^2}}\Big[1-\sigma_1 \frac{q^2}{M_1^2}+\sigma_2\pqty{\frac{q^2}{M_1^2}}^2\Big]},
				\end{equation}
				\item for $f_0(q^2)$, $A_1(q^2)$ and $A_2(q^2)$\cite{PhysRevD.101.013004}:
				\begin{equation}
					F(q^2)=\frac{F(0)}{1-\sigma_1 \frac{q^2}{M_1^2}+\sigma_2\pqty{\frac{q^2}{M_1^2}}^2},
				\end{equation}
			\end{itemlist}
			These parameterizations are not arbitrary but have been tested extensively against experimental observables from semi-leptonic decays, including branching ratios, lepton forward-backward asymmetries, and meson polarization effects. A similar for hadronic decays can be found in Refs. \citen{PhysRevD.56.312, PhysRevD.87.034033, PhysRevD.87.094028}.
			
			The input parameters used in the analysis include the quark compositions of light mesons, their decay constants, and mixing angles that account for flavor \(SU(3)\) breaking effects. For instance, the pseudoscalar mesons \(\eta\) and \(\eta'\) are described in the quark flavor basis with mixing angles determined by phenomenological studies and recent experimental measurements. Such input is crucial for obtaining reliable numerical predictions of the branching fractions.\\
			The theoretical framework described here rests on the assumption that factorization holds to a good approximation in these decays. Even though the factorization method successfully reproduces several observables through empirical adjustments to the number of colors ($N$), capturing the complete decay dynamics will ultimately require the explicit evaluation of final-state interactions and other nonperturbative terms. Recent studies, including those based on QCD factorization\cite{Chai_2022} and perturbative QCD\cite{PhysRevD.76.074018, PhysRevD.103.056006, PhysRevD.85.094003}, have provided additional insights into these subtleties and underscore the need for careful phenomenological analysis in heavy meson decays. In summary, the effective Hamiltonians for charm and bottom meson decays, together with the factorization approximation and well-established input from quark models, form the backbone of the theoretical analysis. This framework enables the calculation of decay amplitudes and branching fractions, while also identifying specific kinematic regimes where nonfactorizable contributions—such as final-state interactions—continue to play a significant role.\newpage
	\section{Amplitudes}
	\begin{table}[h]
		\centering
        \caption{Decay amplitudes of $D$ mesons.\cite{Galkin2023}}
		\begin{tabular}{ll}
			\hline
			\textbf{Reaction} & $\frac{\sqrt{2}}{G_F}\times$\textbf{Amplitude} \\
			\hline
			$D^+ \to \pi^0 \pi^+$ & $ V_{cd}^* V_{ud} \left[ a_1 \mathcal{M}_{D^+ \to \pi^0, \pi^+} + a_2 \mathcal{M}_{D^+ \to \pi^+, \pi^0} \right]$ \\
			$D^+ \to \pi^0 K^+$ & $ V_{cd}^* V_{us} a_1 \mathcal{M}_{D^+ \to \pi^0, K^+}$ \\
			$D^+ \to \eta^{(\prime)} K^+$ & $ V_{cd}^* V_{us} a_1 \mathcal{M}_{D^+ \to \eta^{(\prime)}, K^+}$ \\
			\multirow{2}{*}{$D^+ \to \eta^{(\prime)} \pi^+$} 
			& $\left(V^{*}_{cd}V_{ud}a_{1}\mathcal{M}_{D^{+} \rightarrow\eta^{(\prime)},\pi^{+}} +V^{*}_{cd}V_{ud}a_{2}\mathcal{M}_{D^{+}\rightarrow\pi^{+},\eta^{(\prime)}_{u}} \right.$ \\
			& $\left. +V^{*}_{cs}V_{us}a_{2}\mathcal{M}_{D^{+}\rightarrow\pi^{+},\eta^{(\prime)}_{s}}\right)$ \\
			$D^+ \to \pi^+ \rho^0$ & $ V_{cd}^* V_{ud} \left( a_1 \mathcal{M}_{D^+ \to \rho^0, \pi^+} + a_2 \mathcal{M}_{D^+ \to \pi^+, \rho^0} \right)$ \\
			$D^+ \to \pi^+ \phi$ & $ V_{cs}^* V_{us} a_2 \mathcal{M}_{D^+ \to \pi^+, \phi}$ \\
			$D^+ \to \pi^+ \omega$ & $ V_{cd}^* V_{ud} \left( a_1 \mathcal{M}_{D^+ \to \omega, \pi^+} + a_2 \mathcal{M}_{D^+ \to \pi^+, \omega} \right)$ \\
			$D^+ \to \rho^0 K^+$ & $ V_{cd}^* V_{us} a_1 \mathcal{M}_{D^+ \to \rho^0, K^+}$ \\
			$D^+ \to \rho^+ \phi$ & $ V_{cs}^* V_{us} a_2 \mathcal{M}_{D^+ \to \rho^+, \phi}$ \\
			$D^0 \to K^- \pi^+$ & $ V_{cs}^* V_{ud} a_1 \mathcal{M}_{D^0 \to K^-, \pi^+}$ \\
			$D^0 \to \pi^- \pi^+$ & $ V_{cd}^* V_{ud} a_1 \mathcal{M}_{D^0 \to \pi^-, \pi^+}$ \\
			$D^0 \to \pi^0 \pi^0$ & $2 \times V_{cd}^* V_{ud} a_2 \mathcal{M}_{D^0 \to \pi^0, \pi^0}$ \\
			$D^0 \to K^- K^+$ & $ V_{cs}^* V_{us} a_1 \mathcal{M}_{D^0 \to K^-, K^+}$ \\
			$D^0 \to \eta \eta$ & $2\times \left(V^{*}_{cs}V_{us}a_{2}\mathcal{M}_{D^{0} \rightarrow\eta,\eta_{s}} +V^{*}_{cd}V_{ud}a_{2}\mathcal{M}_{D^{0}\rightarrow\eta,\eta_{u}}\right)$ \\
			$D^0 \to \pi^- K^+$ & $V^{*}_{cd}V_{us}a_{1}\mathcal{M}_{D^{0}\rightarrow\pi^{-},K^{+}}$ \\
			\multirow{2}{*}{$D^0 \to \eta^{(\prime)} \pi^0$} 
			& $\left(V^{*}_{cd}V_{ud}a_{2}\mathcal{M}_{D^{0} \rightarrow\eta^{(\prime)},\pi^{0}} +V^{*}_{cd}V_{ud}a_{2}\mathcal{M}_{D^{0}\rightarrow\pi^{0},\eta^{(\prime)}_{u}} \right.$ \\
			& $\left. +V^{*}_{cs}V_{us}a_{2}\mathcal{M}_{D^{0}\rightarrow\pi^{0},\eta^{(\prime)}_{s}}\right)$ \\
			
			\multirow{2}{*}{$D^0 \to \eta \eta^{\prime}$} 
			& $\left(V^{*}_{cd}V_{ud}a_{2}\mathcal{M}_{D^{0} \rightarrow\eta,\eta^{\prime}_{u}} +V^{*}_{cs}V_{us}a_{2}\mathcal{M}_{D^{0}\rightarrow\eta,\eta^{\prime}_{s}} \right.$ \\
			& $\left. +V^{*}_{cd}V_{ud}a_{2}\mathcal{M}_{D^{0}\rightarrow\eta^{\prime},\eta_{u}} +V^{*}_{cs}V_{us}a_{2}\mathcal{M}_{D^{0}\rightarrow\eta^{\prime},\eta_{s}}\right)$ \\
			
			$D^0 \to \pi^0 \omega$ & $\left(V^{*}_{cd}V_{ud}a_{2}\mathcal{M}_{D^{0} \rightarrow\pi^{0},\omega} +V^{*}_{cd}V_{ud}a_{2}\mathcal{M}_{D^{0}\rightarrow\omega,\pi^{0}}\right)$ \\
			$D^0 \to \eta \omega$ & $\left(V^{*}_{cd}V_{ud}a_{2}\mathcal{M}_{D^{0}\to \eta,\omega} +V^{*}_{cd}V_{ud}a_{2}\mathcal{M}_{D^{0}\to\omega,\eta_{u}} +V^{*}_{cs}V_{us}a_{2}\mathcal{M}_{D^{0}\to\omega,\eta_{s}}\right)$ \\
			$D^0 \to \rho^0 \pi^0$ & $\left(V^{*}_{cd}V_{ud}a_{2}\mathcal{M}_{D^{0}\to \pi^{0},\rho^{0}}+V^{*}_{cd}V_{ud}a_{2}\mathcal{M}_{D^{0}\to\rho^{0},\pi^{0}}\right)$ \\
			$D^0 \to \pi^- \rho^+$ & $V^{*}_{cd}V_{ud}a_{1}\mathcal{M}_{D^{0}\to\pi^{-}, \rho^{+}}$ \\
			$D^0 \to \pi^0 \phi$ & $V^{*}_{cs}V_{us}a_{2}\mathcal{M}_{D^{0}\to\pi^{0}, \phi}$ \\
			$D^0 \to \rho^- \pi^+$ & $V^{*}_{cd}V_{ud}a_{1}\mathcal{M}_{D^{0}\to\rho^{-}, \pi^{+}}$ \\
			$D^0 \to \eta \phi$ & $V^{*}_{cs}V_{us}a_{2}\mathcal{M}_{D^{0}\to\eta,\phi}$ \\
			$D^0 \to K^- \rho^+$ & $V^{*}_{cs}V_{ud}a_{1}\mathcal{M}_{D^{0}\to K^{-},\rho^{ +}}$ \\
			$D^0 \to \eta^{(\prime)} \bar{K}^{*0}$ & $V^{*}_{cs}V_{ud}a_{2}\mathcal{M}_{D^{0}\to\eta^{(\prime)},\bar{K}^{*0}}$ \\
			$D^0 \to \rho^0 \rho^0$ & $2\times V^{*}_{cd}V_{ud}a_{2}\mathcal{M}_{D^{0}\to\rho^{0},\rho^{0}}$ \\
			$D^0 \to \omega \phi$ & $V^{*}_{cs}V_{us}a_{2}\mathcal{M}_{D^{0}\rightarrow\omega,\phi}$ \\
			\hline
		\end{tabular}
	\end{table}
	\newpage
	\begin{table}[h]
		\centering
        \caption{Decay amplitudes of $D_s$ mesons.\cite{Galkin2023}}
		\begin{tabular}{ll}
			\toprule
			\textbf{Reaction} & $\frac{\sqrt{2}}{G_F}\times$\textbf{Amplitude} \\
			\hline
			$D_{s}\to K^{+}\bar{K}^{0}$ &$ V^{*}_{cs}V_{ud}a_{2}\mathcal{M}_{D_{s}\to K^{+},\bar{K}^{0}}$\\
			$D_{s}\rightarrow\eta^{(\prime)}\pi^{+}$&$ V^{*}_{cs}V_{ud}a_{1}\mathcal{M}_{D_{s}\rightarrow\eta^{(\prime)},\pi^{+}}$\\
			$D_{s}\to K^{+}\pi^{0}$ &$ V^{*}_{cd}V_{ud}a_{2}\mathcal{M}_{D_{s}\to K^{+},\pi^{0}}$\\
			$D_{s}\rightarrow\eta^{(\prime)}K^{+}$ &$ \Big{(}V^{*}_{cd}V_{ud}a_{2}\mathcal{M}_{D_{s}\to K^{+},\eta^{(^{\prime})}_{u}} +V^{*}_{cs}V_{us}a_{2}\mathcal{M}_{D_{s}\to K^{+},\eta^{(^{\prime})}_{s}} $\\
			&$ +V^{*}_{cs}V_{us}a_{1}\mathcal{M}_{D_{s}\to\eta^{(\prime)},K^{+}}\Big{)}$\\
			$D_{s}\rightarrow\eta^{(\prime)}\rho^{+})$&$ V^{*}_{cs}V_{ud}a_{1}\mathcal{M}_{D_{s}\to\eta^{(\prime)},\rho^{+}} $\\
			$D_{s}\to K^{+}\omega $&$ V^{*}_{cd}V_{ud}a_{2}\mathcal{M}_{D_{s}\to K^{+},\omega}$\\
			$D_{s}\to K^{+}\rho^{0}$ &$ V^{*}_{cd}V_{ud}a_{2}\mathcal{M}_{D_{s}\to K^{+},\rho^{0}} $\\
			$D_{s}\rightarrow\phi\pi^{+}$ &$ V^{*}_{cs}V_{ud}a_{1}\mathcal{M}_{D_{s}\rightarrow\phi, \pi^{+}} $\\
			$D_{s}\rightarrow\phi\rho^{+}$ &$ V^{*}_{cs}V_{ud}a_{1}\mathcal{M}_{D_{s}\rightarrow\phi, \rho^{+}}$\\
			\botrule
		\end{tabular}
	\end{table}
	\begin{longtable}{ll}
        \caption{\footnotesize Decay amplitudes of $B$ mesons.\cite{Galkin2023}}\\
		  \hline
		\textbf{Reaction} & $\frac{\sqrt{2}}{G_F}\times$\textbf{Amplitude} \\
		\hline
		$B^{+} \to \pi^{+} \eta^{(\prime)}$ & $\Big[ V^{*}_{ub}V_{ud}a_{2} - V^{*}_{tb}V_{td} \Big( 2a_{3} - 2a_{5} - \frac{1}{2}a_{7} + \frac{1}{2}a_{9} + a_{4}  $ \\ 
		& $-\frac{1}{2}a_{10} + \frac{m_{\eta^{(\prime)}}^{2}}{m_{s}(m_{b}-m_{d})} \big( a_{6} - \frac{1}{2}a_{8} \big) \Big( \frac{f^{s}_{\eta^{(\prime)}}}{{f}^{9
			u}_{\eta^{(\prime)}}} - 1 \Big) r_{\eta^{(\prime)}} \Big) \Big] \mathcal{M}_{B^{+} \to \pi^{+}, \eta^{(\prime)}_{u}}  $\\
		& $- V^{*}_{tb}V_{td} \Big( a_{3} - a_{5} + \frac{1}{2}a_{7} - \frac{1}{2}a_{9} \Big) \mathcal{M}_{B^{+} \to \pi^{+}, \eta^{(\prime)}_{s}} + \Big[ V^{*}_{ub}V_{ud}a_{1} $\\
		& $-V^{*}_{tb}V_{td} \Big( a_{4} + a_{10} + \frac{2m_{\pi^{+}}^{2}}{(m_{u}+m_{d})(m_{b}-m_{u})} (a_{6} + a_{8}) \Big) \Big] \mathcal{M}_{B^{+} \to \eta^{(\prime)}, \pi^{+}} $ \\
		$B^{+} \to \pi^{+} \omega$ & $ \Big[ V^{*}_{ub}V_{ud}a_{2} - V^{*}_{tb}V_{td} \Big( 2a_{3} + 2a_{5} + \frac{1}{2}a_{7} + \frac{1}{2}a_{9} + a_{4}$\\
		& $- \frac{1}{2}a_{10} \Big) \Big] \mathcal{M}_{B^{+} \to \pi^{+}, \omega} + \Big[V^{*}_{ub}V_{ud}a_{1} - V^{*}_{tb}V_{td} \Big( a_{4} + a_{10}$\\
		& $- \frac{2m_{\pi^{+}}^{2}}{(m_{u}+m_{d})(m_{b}+m_{u})} (a_{6} + a_{8}) \Big) \Big] \mathcal{M}_{B^{+} \to \omega, \pi^{+}} $\\
		$B^{+} \to \rho^{+} \eta^{(\prime)}$ & $\Big[ V^{*}_{ub}V_{ud}a_{2} - V^{*}_{tb}V_{td} \Big( 2a_{3} - 2a_{5} - \frac{1}{2}a_{7} + \frac{1}{2}a_{9} + a_{4}$ \\
		& $- \frac{1}{2}a_{10} - \frac{m_{\eta^{(\prime)}}^{2}}{m_{s}(m_{b}+m_{d})} \big(a_{6} - \frac{1}{2}a_{8} \big) \Big( \frac{f^{s}_{\eta^{(\prime)}}}{f^{u}_{\eta^{(\prime)}}} - 1 \Big) r_{\eta^{(\prime)}} \Big) \Big] \mathcal{M}_{B^{+} \to \rho^{+}, \eta^{(\prime)}_{u}} $ \\
		& $- V^{*}_{tb}V_{td} \Big( a_{3} - a_{5} + \frac{1}{2}a_{7} - \frac{1}{2}a_{9} \Big) \mathcal{M}_{B^{+} \to \rho^{+}, \eta^{(\prime)}_{s}}  $\\
		&$ + \Big[ V^{*}_{ub}V_{ud}a_{1} - V^{*}_{tb}V_{td} (a_{4} + a_{10}) \Big] \mathcal{M}_{B^{+} \to \eta^{(\prime)}, \rho^{+}}$ \\
		$B^+ \to \pi^+ K^0$ & $ -V_{tb}^* V_{ts} \Big( a_4 - \frac{1}{2} a_{10}+ \frac{2m_{K^0}^2}{(m_s + m_d)(m_b - m_d)} (a_6 - \frac{1}{2} a_8) \Big) \mathcal{M}_{B^+ \to \pi^+, K^0}$ \\
		$B^+ \to \rho^+ K^0)$ & $ - V_{tb}^* V_{ts} \Big( a_4 - \frac{1}{2} a_{10} - \frac{2m_{K^0}^2}{(m_s + m_d)(m_b + m_d)} (a_6 - \frac{1}{2} a_8) \Big) \mathcal{M}_{B^+ \to \rho^+, K^0} $ \\
		$B^+ \to \pi^+ \pi^0)$ & $ \Big[ V_{ub}^* V_{ud} a_2 - V_{tb}^* V_{td} \Big( \frac{3}{2} a_9 - \frac{3}{2} a_7$\\
		&$ - a_4 + \frac{1}{2} a_{10} - \frac{m_{\pi^0}^2}{m_d(m_b - m_d)} \Big( a_6 - \frac{1}{2} a_8 \Big) \Big) \Big] \mathcal{M}_{B^+ \to \pi^+, \pi^0} $\\
		&$ + \Big[ V_{ub}^* V_{ud} a_1 - V_{tb}^* V_{td} \Big( a_4 + a_{10} $\\
		&$ + \frac{2m_{\pi^+}^2}{(m_u + m_d)(m_b - m_u)} (a_6 + a_8) \Big) \Big] \mathcal{M}_{B^+ \to \pi^0, \pi^+}, $ \\
		$B^+ \to \pi^+ \rho^0$ & $\Big[ V_{ub}^* V_{ud} a_2 - V_{tb}^* V_{td} \Big( -a_4 + \frac{1}{2} a_{10} + \frac{3}{2} a_9 + \frac{3}{2} a_7 \Big) \Big] \mathcal{M}_{B^+ \to \pi^+, \rho^0}$ \\
		&$ + \Big[ V_{ub}^* V_{ud} a_1 - V_{tb}^* V_{td} \Big( a_4 + a_{10}$\\
		&$ - \frac{2m_{\pi^+}^2}{(m_u + m_d)(m_b + m_u)} (a_6 + a_8) \Big] \mathcal{M}_{B^+ \to \rho^0, \pi^+}$ \\
		$B^+ \to \rho^+ \pi^0$ & $ \Big[ V_{ub}^* V_{ud} a_2 - V_{tb}^* V_{td} \Big( \frac{3}{2} a_9 - \frac{3}{2} a_7 - a_4 + \frac{1}{2} a_{10} $\\
		&$ + \frac{m_{\pi^0}^2}{m_d(m_b + m_d)} \Big( a_6 - \frac{1}{2} a_8 \Big) \Big) \Big] \mathcal{M}_{B^+ \to \rho^+, \pi^0}$\\
		&$ + \Big[ V_{ub}^* V_{ud} a_1 - V_{tb}^* V_{td} \Big( a_4 + a_{10} \Big) \Big] \mathcal{M}_{B^+ \to \pi^0, \rho^+} $ \\
		$B^+ \rightarrow \pi^+ \phi$ & $ - V_{tb}^* V_{td} \left( a_3 + a_5 - \frac{1}{2} a_7 - \frac{1}{2} a_9 \right) \mathcal{M}_{B^+ \rightarrow \pi^+, \phi}, $ \\
		$B^+ \rightarrow \rho^+ \rho^0)$ & $\Big[ V_{ub}^* V_{ud} a_2 - V_{tb}^* V_{td} \Big( \frac{3}{2} a_9 + \frac{3}{2} a_7 - a_4 + \frac{1}{2} a_{10} \Big) \Big] \mathcal{M}_{B^+ \rightarrow \rho^+,\rho^0}$\\
		&$ + \Big[ V_{ub}^* V_{ud} a_1 - V_{tb}^* V_{td} \Big( a_4 + a_{10} \Big) \Big] \mathcal{M}_{B^+ \rightarrow \rho^0,\rho^+}$ \\
		$B^+ \rightarrow \rho^+ \omega$ & $ \Big[ V_{ub}^* V_{ud} a_2 - V_{tb}^* V_{td} \Big( 2a_3 + 2a_5 + \frac{1}{2} a_7 + \frac{1}{2} a_9 + a_4 $\\
		&$ - \frac{1}{2} a_{10} \Big) \Big] \mathcal{M}_{B^+ \rightarrow \rho^+,\omega}  + \Big[ V_{ub}^* V_{ud} a_1 $\\
		&$- V_{tb}^* V_{td} \Big( a_4 + a_{10} \Big) \Big] \mathcal{M}_{B^+ \rightarrow \omega,\rho^+} $ \\
		$B^0 \rightarrow D^- \pi^+$ & $V_{cb}^* V_{ud} a_1 \mathcal{M}_{B^0 \rightarrow D^-,\pi^+}$ \\
		$B^0 \rightarrow D^- K^+$ & $V_{cb}^* V_{us} a_1 \mathcal{M}_{B^0 \rightarrow D^-,K^+}$ \\
		$B^0 \rightarrow \pi^- K^+$ & $ \Big[ V_{ub}^* V_{us} a_1 - V_{tb}^* V_{ts} \Big( a_4 + a_{10} $ \\
		&$+ \frac{2m_K^2}{(m_u + m_s)(m_b - m_u)} (a_6 + a_8) \Big) \Big] \mathcal{M}_{B^0 \rightarrow \pi^-, K^+}$\\
		$B^0 \rightarrow \pi^- \pi^+$ & $ \Big[ V_{ub}^* V_{ud} a_1 - V_{tb}^* V_{td} \Big( a_4 + a_{10}$\\
		&$+ \frac{2m_{\pi^+}^2}{(m_u + m_d)(m_b - m_u)} (a_6 + a_8) \Big) \Big] \mathcal{M}_{B^0 \rightarrow \pi^-, \pi^+}$ \\
		$B^0 \rightarrow \pi^0 \pi^0$ & $ 2\times \Big[ V_{ub}^* V_{ud} a_2 - V_{tb}^* V_{td} \Big( \frac{3}{2} a_9 - \frac{3}{2} a_7 - a_4 + \frac{1}{2} a_{10}$\\
		&$ - \frac{m_{\pi^0}^2}{m_d(m_b - m_d)} \Big( a_6 - \frac{1}{2} a_8 \Big) \Big) \Big] \mathcal{M}_{B^0 \rightarrow \pi^0, \pi^0}$ \\
		$B^{0} \to \pi^{0} \eta^{(\prime)}$ & $\Big[ V^{*}_{ub}V_{ud}a_{2} - V^{*}_{tb} V_{td} \Big( 2a_{3} - 2a_{5} - \frac{1}{2}a_{7} + \frac{1}{2}a_{9} + a_{4} - \frac{1}{2}a_{10} $\\
		&$ + \frac{m_{\eta^{(\prime)}}^{2}}{m_{s}(m_{b}-m_{d})} \Big( a_{6} - \frac{1}{2}a_{8} \Big) \Big( \frac{f_{{\eta}^{(\prime)}}^{s}}{f_{{\eta}^{(\prime)}}^{u}} - 1 \Big) r_{\eta^{(\prime)}} \Big) \Big] \mathcal{M}_{B^{0} \to \pi^{0},\eta_{u}^{(\prime)}} $\\
		&$ - V^{*}_{tb}V_{td} \Big( a_{3} - a_{5} + \frac{1}{2}a_{7} - \frac{1}{2}a_{9} \Big) \mathcal{M}_{B^{0} \to \pi^{0},\eta_{s}^{(\prime)}} $\\
		&$ + \Big[ V^{*}_{ub}V_{ud}a_{2} - V^{*}_{tb}V_{td} \Big( \frac{3}{2}a_{9} - \frac{3}{2}a_{7} - a_{4} + \frac{1}{2}a_{10} $\\
		&$ - \frac{m_{\pi^{0}}^{2}}{m_{d}(m_{b}-m_{d})} \Big( a_{6} - \frac{1}{2}a_{8} \Big) \Big) \Big] \mathcal{M}_{B^{0} \to \eta^{(\prime)},\pi^{0}}$\\
		$B^{0} \to \eta\eta$ & $ 2\times \Big[ V^{*}_{ub}V_{ud}a_{2} - V^{*}_{tb}V_{td} \Big( 2a_{3} - 2a_{5} - \frac{1}{2}a_{7} + \frac{1}{2}a_{9} + a_{4} - \frac{1}{2}a_{10}$\\
		&$ + \frac{2m_{\eta}^{2}}{(m_{s}+m_{s})(m_{b}-m_{d})} \Big( a_{6} - \frac{1}{2}a_{8} \Big) \Big( \frac{f_{\eta}^{s}}{f_{\eta}^{u}} - 1 \Big) r_{\eta} \Big) \Big] \mathcal{M}_{B^{0} \to \eta,\eta_{u}}$\\
		&$ - V^{*}_{tb}V_{td} \Big( a_{3} - a_{5} + \frac{1}{2}a_{7} - \frac{1}{2}a_{9} \Big) \mathcal{M}_{B^{0} \to \eta,\eta_{s}}$ \\
		$B^0 \rightarrow \eta^{\prime} \eta^{\prime}$ & $ 2\times\Big[ V_{ub}^*V_{ud}a_2 - V_{tb}^*V_{td} \Big( 2a_3 - 2a_5 - \frac{1}{2}a_7 + \frac{1}{2}a_9 + a_4 - \frac{1}{2}a_{10}$\\
		&$ + \frac{m_{\eta^{\prime}}^2}{m_s(m_b - m_d)} \Big( a_6 - \frac{1}{2}a_8 \Big) \Big( \frac{f_{\eta^{\prime}}^s}{f_{\eta^{\prime}}^u} - 1 \Big) r_{\eta^{\prime}} \Big) \Big] \mathcal{M}_{B^0 \rightarrow \eta^{\prime}, \eta_u}$\\
		&$ - V_{tb}^*V_{td} \Big( a_3 - a_5 + \frac{1}{2}a_7 - \frac{1}{2}a_9 \Big) \mathcal{M}_{B^0 \rightarrow \eta^{\prime}, \eta_s}$\\
		$B^0 \rightarrow D^{*-}\pi^+$ & $ V_{cb}^*V_{ud}a_1 \mathcal{M}_{B^0 \rightarrow D^{*-}\pi^+}$\\
		$B^0 \rightarrow D^{-}\rho^+$ & $ V_{cb}^*V_{ud}a_1 \mathcal{M}_{B^0 \rightarrow D^{-}\rho^+}$\\
		$B^0 \rightarrow \rho^0 \pi^0$ & $ V_{ub}^*V_{ud}a_2 - V_{tb}^*V_{td} \Big( \frac{3}{2}a_7 + \frac{3}{2}a_9 - a_4 + \frac{1}{2}a_{10} \Big) \Big\} \mathcal{M}_{B^0 \rightarrow \pi^0, \rho^0}$\\
		&$ + \Big[ V_{ub}^*V_{ud}a_2 - V_{tb}^*V_{td} \Big( \frac{3}{2}a_9 - \frac{3}{2}a_7 - a_4 + \frac{1}{2}a_{10}$\\
		&$ + \frac{m_{\pi^0}^2}{m_d(m_b + m_d)} \Big( a_6 - \frac{1}{2}a_8 \Big) \Big) \Big] \mathcal{M}_{B^0 \rightarrow \rho^0, \pi^0}$\\
		$B^0 \rightarrow \omega \pi^0$ & $\Big[ V_{ub}^* V_{ud} a_2 - V_{tb}^* V_{td} \Big( 2a_3 + 2a_5 + \frac{1}{2} a_7 + \frac{1}{2} a_9 + a_4 $\\
		&$ - \frac{1}{2} a_{10} \Big) \Big] \mathcal{M}_{B^0 \rightarrow \pi^0, \omega} + \Big[ V_{ub}^* V_{ud} a_2 - V_{tb}^* V_{td} \Big( \frac{3}{2} a_9 - \frac{3}{2} a_7 - a_4 $\\
		&$+ \frac{1}{2} a_{10} + \frac{m_{\pi^0}^2}{m_d(m_b + m_d)} \Big( a_6 - \frac{1}{2} a_8 \Big) \Big) \Big] \mathcal{M}_{B^0 \rightarrow \omega, \pi^0}$ \\
		$B^0 \rightarrow \rho^- K^+$ & $ \Big[ V_{ub}^* V_{us} a_1 - V_{tb}^* V_{ts} \Big( a_4 + a_{10}$\\
		&$- \frac{2m_{K^+}^2}{(m_s + m_u)(m_b + m_u)} \Big( a_6 + a_8 \Big) \Big) \Big] \mathcal{M}_{B^0 \rightarrow \rho^-, K^+}$ \\
		$B^0 \rightarrow \pi^- K^{*+}$ & $\Big[ V_{ub}^* V_{us} a_1 - V_{tb}^* V_{ts} \Big( a_4 + a_{10} \Big) \Big] \mathcal{M}_{B^0 \rightarrow \pi^-, K^{*+}} $ \\
		$B^0 \rightarrow \rho^- \pi^+$ & $\Big[ V_{ub}^* V_{ud} a_1 - V_{tb}^* V_{td} \Big( a_4 + a_{10} $\\
		&$ - \frac{2m_{\pi^+}^2}{(m_u + m_d)(m_b + m_u)} ( a_6 + a_8) \Big) \Big] \mathcal{M}_{B^0 \rightarrow \rho^-, \pi^+} $ \\
		$B^0 \rightarrow D^- K^{*+}$ & $V_{cb}^* V_{us} a_1 \mathcal{M}_{B^0 \rightarrow D^-, K^{*+}}$\\
		$B^0 \rightarrow D^{*-} K^+$ & $V_{cb}^* V_{us} a_1 \mathcal{M}_{B^0 \rightarrow D^{*-}, K^+}$ \\
		$B^0 \rightarrow \eta^{\prime} \eta$ & $\Big[ V_{ub}^* V_{ud} a_2 - V_{tb}^* V_{td} \Big( 2a_3 - 2a_5 - \frac{1}{2} a_7 + \frac{1}{2} a_9 + a_4 - \frac{1}{2} a_{10} $\\
		&$ + \frac{m_{\eta}^{2}}{m_s(m_b - m_d)} \Big( a_6 - \frac{1}{2} a_8 \Big) \Big( \frac{f_\eta^s}{f_\eta^u} - 1 \Big) r_\eta \Big) \Big] \mathcal{M}_{B^0 \rightarrow \eta^{\prime}, \eta_u} $\\
		&$ - V_{tb}^* V_{td} \Big( a_3 - a_5 + \frac{1}{2} a_7 - \frac{1}{2} a_9 \Big) \mathcal{M}_{B^0 \rightarrow \eta^{\prime}, \eta_s} $\\
		&$ + \Big[ V_{ub}^* V_{ud} a_2 - V_{tb}^* V_{td} \Big( 2a_3 - 2a_5 - \frac{1}{2} a_7 + \frac{1}{2} a_9 + a_4 - \frac{1}{2} a_{10} $\\
		&$ + \frac{m_{\eta'}^2}{m_s(m_b - m_d)} \Big( a_6 - \frac{1}{2} a_8 \Big) \Big( \frac{f_{\eta^{\prime}}^s}{f_{\eta^{\prime}}^u} - 1 \Big) r_{\eta^{\prime}} \Big) \Big] \mathcal{M}_{B^0 \rightarrow \eta, \eta_u^{\prime}} $\\
		&$ - V_{tb}^* V_{td} \Big( a_3 - a_5 + \frac{1}{2} a_7 - \frac{1}{2} a_9 \Big) \mathcal{M}_{B^0 \rightarrow \eta, \eta_s^{\prime}}$ \\
		$B^0 \to \rho^0\eta^{(\prime)}$ & $\Big[ V_{ub}^*V_{ud}a_2 - V_{tb}^*V_{td} \Big( 2a_3 - 2a_5 - \frac{1}{2}a_7 + \frac{1}{2}a_9 + a_4 - \frac{1}{2}a_{10}$\\
		&$ - \frac{m_{\eta^{(\prime)}}^2}{m_s(m_b + m_d)} \Big( a_6 - \frac{1}{2}a_8 \Big) \Big( \frac{f_{\eta^{(\prime)}}^s}{f_{\eta^{(\prime)}}^u} - 1 \Big) r_{\eta^{(\prime)}} \Big) \Big] \mathcal{M}_{B^0 \to \rho^0, \eta_u^{(\prime)}}$\\
		&$ - V_{tb}^*V_{td} \Big( a_3 - a_5 + \frac{1}{2}a_7 - \frac{1}{2}a_9 \Big) \mathcal{M}_{B^0 \to \rho^0, \eta_s^{(\prime)}}$\\
		&$ + \Big[ V_{ub}^*V_{ud}a_2 - V_{tb}^*V_{td} \Big( \frac{3}{2}a_7 + \frac{3}{2}a_9 - a_4 + \frac{1}{2}a_{10} \Big) \Big] \mathcal{M}_{B^0 \to \eta^{(\prime)}, \rho^0}$ \\
		$B^0 \rightarrow \omega \eta^{(\prime)}$ & $\Big[ V_{ub}^* V_{ud} a_2 - V_{tb}^* V_{td} \Big( 2a_3 - 2a_5 - \frac{1}{2} a_7 + \frac{1}{2} a_9 + a_4 - \frac{1}{2} a_{10}  $\\
		&$ - \frac{m_{\eta^{(\prime)}}^2}{m_d(m_b + m_d)} \Big( a_6 - \frac{1}{2} a_8 \Big) \Big( \frac{f_{\eta^{(\prime)}}^s}{f_{\eta^{(\prime)}}^u} - 1 \Big) r_{\eta^{(\prime)}} \Big) \Big] \mathcal{M}_{B^0 \rightarrow \omega, \eta^{(\prime)}_u} $ \\
		&$ - V_{tb}^* V_{td} \Big( a_3 - a_5 + \frac{1}{2} a_7 - \frac{1}{2} a_9 \Big) \mathcal{M}_{B^0 \rightarrow \omega, \eta^{(\prime)}_s} + \Big[ V_{ub}^* V_{ud} a_2$ \\
		&$- V_{tb}^* V_{td} \Big( 2a_3 + 2a_5 + \frac{1}{2} a_7 + \frac{1}{2} a_9 + a_4 - \frac{1}{2} a_{10} \Big) \Big] \mathcal{M}_{B^0 \rightarrow \eta^{(\prime)}, \omega}$\\
		$B^0 \rightarrow \pi^- \rho^+$ & $ \Big[ V_{ub}^* V_{ud} a_1 - V_{tb}^* V_{td} \Big( a_4 + a_{10} \Big) \Big] \mathcal{M}_{B^0 \rightarrow \pi^-, \rho^+} $ \\
		$B^0 \rightarrow \rho^- \rho^+$ & $ \Big[ V_{ub}^* V_{ud} a_1 - V_{tb}^* V_{td} \Big( a_4 + a_{10} \Big) \Big] \mathcal{M}_{B^0 \rightarrow \rho^-, \rho^+} $ \\
		$B^0 \rightarrow \rho^0 \rho^0$ & $ 2\times \Big[ V_{ub}^* V_{ud} a_2 - V_{tb}^* V_{td} \Big( \frac{3}{2} a_9 + \frac{3}{2} a_7 - a_4 + \frac{1}{2} a_{10} \Big) \Big] \mathcal{M}_{B^0 \rightarrow \rho^0, \rho^0}$ \\
		$B^0 \rightarrow \omega \omega$ & $ 2 \times \Big[ V_{ub}^* V_{ud} a_2 - V_{tb}^* V_{td} \Big( 2 a_3 + 2 a_5 + \frac{1}{2} a_7 + \frac{1}{2} a_9 $ \\
		&$ + a_4 - \frac{1}{2} a_{10} \Big) \Big] \mathcal{M}_{B^0 \rightarrow \omega, \omega}$\\
		$B^0 \rightarrow \omega \rho^0$ & $\Big[ V_{ub}^* V_{ud} a_2 - V_{tb}^* V_{td} \Big( 2a_3 + 2a_5 + \frac{1}{2} a_7 + \frac{1}{2} a_9 + a_4 $\\
		&$ - \frac{1}{2} a_{10} \Big) \Big] \mathcal{M}_{B^0 \rightarrow \rho^0, \omega} + \Big[ V_{ub}^* V_{ud} a_2- V_{tb}^* V_{td} \Big( \frac{3}{2} a_7 + \frac{3}{2} a_9  $\\
		&$ - a_4 + \frac{1}{2} a_{10} \Big) \Big] \mathcal{M}_{B^0 \rightarrow \omega, \rho^0}$ \\
		$B^0 \rightarrow D^{*-} \rho^{+}$ & $V_{cb}^{*} V_{ud} a_1 \mathcal{M}_{B^0 \rightarrow D^{*-}, \rho^{+}}$ \\
        $B_s^0 \to D^{*-} K^{*+}$ & $V_{cb}^* V_{us} a_1 \mathcal{M}_{B_s^0 \to D^{*-} K^{*+}}$ \\
		\hline
	\end{longtable}
	\begin{table}[!h]
		\centering
        \caption{Decay amplitudes of $B_s$ mesons.\cite{Galkin2023}}
		\begin{tabular}{ll}
			\hline
			\textbf{Reaction} & $\frac{\sqrt{2}}{G_F}\times$\textbf{Amplitude} \\
			\hline
			$B_s \to D_s^- \pi^+$ & $V_{cb}^* V_{ud} a_1 \mathcal{M}_{B_s \to D_s^- \pi^+}$ \\
			$B_s \to D_s^- K^+$ & $V_{cb}^* V_{us} a_1 \mathcal{M}_{B_s \to D_s^- K^+} $ \\
			$B_s \to \pi^+ K^-$ & $\Big[ V_{ub}^* V_{ud} a_1 - V_{tb}^* V_{td} \Big( a_4 + a_{10}$\\
			&$ + \frac{2m_{\pi^+}^2}{(m_u + m_d)(m_b - m_u)} \Big( a_6 + a_8 \Big) \Big) \Big] \mathcal{M}_{B_s \to K^-, \pi^+}$\\
			$B_s \to K^+ K^-$ & $\Big[ V_{ub}^* V_{us} a_1 - V_{tb}^* V_{ts} \Big( a_4 + a_{10}$\\
			&$ + \frac{2m_{K^+}^2}{(m_u + m_s)(m_b - m_u)} \Big( a_6 + a_8 \Big) \Big) \Big] \mathcal{M}_{B_s \to K^-, K^+}$\\
			$B_s \to D_s^- \rho^+$ & $V_{cb}^* V_{ud} a_1 \mathcal{M}_{B_s \to D_s^- \rho^+}$\\
			$B_s \to D_s^{*-} \pi^+$ & $ V_{cb}^* V_{ud} a_1 \mathcal{M}_{B_s \to D_s^{*-} \pi^+}$\\
			$B_s \to D_s^{*-} K^+$ & $V_{cb}^* V_{us} a_1 \mathcal{M}_{B_s \to D_s^{*-} K^+}$\\
			$B_s \to D_s^- K^{*+}$ & $V_{cb}^* V_{us} a_1 \mathcal{M}_{B_s \to D_s^- K^{*+}}$ \\
			$B_s \to K^{*-} \pi^+$ & $\Big[ V_{ub}^* V_{ud} a_1 - V_{tb}^* V_{td} \Big( a_4 + a_{10}$\\
			&$ - \frac{2m_{\pi^+}^2}{(m_u + m_d)(m_b + m_u)} \Big( a_6 + a_8 \Big) \Big) \Big] \mathcal{M}_{B_s \to K^{*-}, \pi^+}$\\
			$B_s \to K^{*-} K^+$ & $\Big[ V_{ub}^* V_{us} a_1 - V_{tb}^* V_{ts} \Big( a_4 + a_{10}$\\
			&$ - \frac{2m_{K^+}^2}{(m_u + m_s)(m_b + m_u)} \Big( a_6 + a_8 \Big) \Big) \Big] \mathcal{M}_{B_s \to K^{*-}, K^+}$ \\
			$B_s \to K^- K^{*+}$ & $\Big[ V_{ub}^* V_{us} a_1 - V_{tb}^* V_{ts} (a_4 + a_{10}) \Big] \mathcal{M}_{B_s \to K^-, K^{*+}} $ \\
			$B_s \to D_s^{*-} K^{*+}$ & $V_{cb}^* V_{us} a_1 \mathcal{M}_{B_s \to D_s^{*-} K^{*+}} $ \\
			$B_s \to D_s^{*-} \rho^+$ & $V_{cb}^* V_{ud} a_1 \mathcal{M}_{B_s \to D_s^{*-} \rho^+}$\\
			\hline
		\end{tabular}
	\end{table}
	\newpage	
	\section{Results and Discussion}
	\begin{table}[h]
		\tbl{Branching fractions for $D$ decays and their comparison with other references\label{DBfs}}{\begin{tabular}{lccccccc}
		\toprule
		\multirow{3}[5]{*}{Channel} & \multicolumn{4}{c}{(This Work)  } & \multirow{3}[5]{*}{PDG \cite{PDG2024}} & \multirow{3}[5]{*}{Ref.\citen{Galkin2023}} & \multirow{3}[5]{*}{Ref.\citen{PhysRevD.92.014032}} \\
		\cmidrule{2-5}      & \multicolumn{2}{c}{Hydrogenic} & \multicolumn{2}{c}{Gaussian} &       &       &  \\
		\cmidrule{2-5}      & $N=3$ & $N\to\infty$ & $N=3$ & $N\to\infty$ &       &       &  \\
		\midrule
		$D^+ \to \pi^0 \pi^+$ & $3.44   \times 10^{-3}$ & $1.91 \times10^{-3}$ & $4.55   \times 10^{-3}$ & $2.53\times10^{-3}$ & $(1.247 \pm 0.033)\times10^{-3}$ & $1.30\times10^{-3}$ & $(8.89 \pm 4.51) \times10^{-3}$ \\
		$D^+ \to \pi^0 K^+$ & $2.77 \times 10^{-4}$ & $3.61\times10^{-4}$ & $3.72 \times 10^{-4}$ & $4.84\times10^{-4}$ & $(2.08 \pm 0.21)\times10^{-4}$ & $2.52\times10^{-4}$ & $(3.07 \pm 1.02)\times10^{-4}$ \\
		$D^+ \to \eta K^+$ & $1.55 \times 10^{-4}$ & $2.01\times10^{-4}$ & $2.22 \times 10^{-4}$ & $2.89\times10^{-4}$ & $(1.25 \pm 0.16)\times10^{-4}$ & $3.01\times10^{-4}$ & $(0.98 \pm 0.26)\times10^{-4}$ \\
		$D^+ \to \eta^{\prime} K^+$ & $0.48 \times 10^{-4}$ & $0.62\times10^{-4}$ & $0.88 \times 10^{-4}$ & $1.15\times10^{-4}$ & $(1.85 \pm 0.2)\times10^{-4}$ & $1.21\times10^{-4}$ & $(1.4 \pm 0.39)\times10^{-4}$ \\
		$D^+ \to \eta \pi^+$ & $1.59 \times 10^{-3}$ & $0.19\times10^{-3}$ & $2.23 \times 10^{-3}$ & $0.30\times10^{-3}$ & $(3.77 \pm 0.09)\times10^{-3}$ & $0.31\times10^{-3}$ & $(4.72 \pm 0.21)\times10^{-3}$ \\
		$D^+ \to \eta^{\prime} \pi^+$ & $1.06 \times 10^{-3}$ & $2.22\times10^{-3}$ & $1.74 \times 10^{-3}$ & $3.55\times10^{-3}$ & $(4.97 \pm 0.19)\times10^{-3}$ & $3.66\times10^{-3}$ & $(6.76 \pm 2.19)\times10^{-3}$ \\
		$D^+ \to \pi^+ \rho^0$ & $18.6 \times 10^{-4}$ & $3.13\times10^{-4}$ & $28.2 \times 10^{-4}$ & $4.72\times10^{-4}$ & $(8.4 \pm 0.8)\times10^{-4}$ & $2.43\times10^{-4}$ & - \\
		$D^+ \to \pi^+ \phi$ & $0.04 \times 10^{-3}$ & $1.22\times10^{-3}$ & $0.06 \times 10^{-3}$ & $2.16\times10^{-3}$ & $(5.7 \pm 0.14)\times10^{-3}$ & $2.23\times10^{-3}$ & - \\
		$D^+ \to \pi^+ \omega$ & $15.1 \times 10^{-4}$ & $2.45\times10^{-4}$ & $22.9 \times 10^{-4}$ & $3.72\times10^{-4}$ & $(2.8 \pm 0.6)\times10^{-4}$ & $1.91\times10^{-4}$ & - \\
		$D^+ \to \rho^0 K^+$ & $1.38 \times 10^{-4}$ & $1.79\times10^{-4}$ & $2.42 \times 10^{-4}$ & $3.15\times10^{-4}$ & $(1.9 \pm 0.5)\times10^{-4}$ & $1.65\times10^{-4}$ & - \\
		$D^+ \to \rho^+ \phi$ & $0.008 \times 10^{-2}$ & $0.25\times10^{-2}$ & $0.008 \times 10^{-2}$ & $0.26\times10^{-2}$ & $<1.5\times10^{-2}$ & $0.27\times10^{-2}$ & - \\
		$D^0 \to K^- \pi^+$ & $2.96 \times 10^{-2}$ & $3.96\times10^{-2}$ & $4.08 \times 10^{-2}$ & $5.46\times10^{-2}$ & $(3.94 \pm 0.03)\times10^{-2}$ & $5.44\times10^{-2}$ & $(3.70 \pm 1.33)\times10^{-2}$ \\
		$D^0 \to \pi^- \pi^+$ & $1.62 \times 10^{-3}$ & $2.14\times10^{-3}$ & $2.14 \times 10^{-3}$ & $2.83\times10^{-3}$ & $(1.454 \pm 0.024)\times10^{-3}$ & $2.86\times10^{-3}$ & $(1.44 \pm 0.027)\times10^{-3}$ \\
		$D^0 \to \pi^0 \pi^0$ & $0.44 \times 10^{-4}$ & $14.0\times10^{-4}$ & $0.58 \times 10^{-4}$ & $18.5\times10^{-4}$ & $(8.26 \pm 0.25)\times10^{-4}$ & $2.35\times10^{-4}$ & $(1.14 \pm 0.56)\times10^{-3}$ \\
		$D^0 \to K^- K^+$ & $2.10 \times 10^{-3}$ & $2.78\times10^{-3}$ & $2.97 \times 10^{-3}$ & $3.92\times10^{-3}$ & $(4.08 \pm 0.06)\times10^{-3}$ & $3.96\times10^{-3}$ & $(4.06 \pm 0.77)\times10^{-3}$ \\
		$D^0 \to \eta \eta$ & $0.04 \times 10^{-3}$ & $1.42\times10^{-3}$ & $0.06 \times 10^{-3}$ & $2.05\times10^{-3}$ & $(2.11 \pm 0.19)\times10^{-3}$ & $2.07\times10^{-3}$ & $(1.27 \pm 0.27)\times10^{-3}$ \\
		$D^0 \to \pi^- K^+$ & $1.10 \times 10^{-4}$ & $1.43\times10^{-4}$ & $1.47 \times 10^{-4}$ & $1.92\times10^{-4}$ & $(1.5 \pm 0.07)\times10^{-4}$ & $1.97\times10^{-4}$ & $(1.77 \pm 0.88)\times10^{-4}$ \\
		$D^0 \to \eta \pi^0$ & $0.47 \times 10^{-4}$ & $15.1\times10^{-4}$ & $0.65 \times 10^{-4}$ & $20.7\times10^{-4}$ & $(6.3 \pm 0.6)\times10^{-4}$ & $0.85\times10^{-4}$ & $(1.47 \pm 0.9)\times10^{-3}$ \\
		$D^0 \to \eta^{\prime} \pi^0$ & $0.0004 \times 10^{-4}$ & $0.01\times10^{-4}$ & $0.001 \times 10^{-4}$ & $0.05\times10^{-4}$ & $(9.2 \pm 1.0)\times10^{-4}$ & $2.2\times10^{-4}$ & $(2.17 \pm 0.65)\times10^{-3}$ \\
		$D^0 \to \eta \eta^{\prime}$ & $0.0007 \times 10^{-3}$ & $0.02\times10^{-3}$ & $0.001 \times 10^{-3}$ & $0.05\times10^{-3}$ & $(1.01 \pm 0.19)\times10^{-3}$ & $0.05\times10^{-3}$ & $(9.53 \pm 1.83)\times10^{-4}$ \\
		$D^0 \to \pi^0 \omega$ & $0.41 \times 10^{-4}$ & $13.2\times10^{-4}$ & $0.63 \times 10^{-4}$ & $19.9\times10^{-4}$ & $(1.17 \pm 0.35)\times10^{-4}$ & $0.36\times10^{-4}$ & - \\
		$D^0 \to \eta \omega$ & $0.03 \times 10^{-3}$ & $0.94\times10^{-3}$ & $0.05 \times 10^{-3}$ & $1.77\times10^{-3}$ & $(1.98 \pm 0.18)\times10^{-3}$ & $0.89\times10^{-3}$ & - \\
		$D^0 \to \rho^0 \pi^0$ & $0.05 \times 10^{-3}$ & $1.60\times10^{-3}$ & $0.07 \times 10^{-3}$ & $2.42\times10^{-3}$ & $(3.86 \pm 0.24)\times10^{-3}$ & $0.61\times10^{-3}$ & - \\
		$D^0 \to \pi^- \rho^+$ & $0.29 \times 10^{-2}$ & $0.40\times10^{-2}$ & $0.44 \times 10^{-2}$ & $0.58\times10^{-2}$ & $(1.01 \pm 0.05)\times10^{-2}$ & $0.59\times10^{-2}$ & - \\
		$D^0 \to \pi^0 \phi$ & $0.01 \times 10^{-3}$ & $0.49\times10^{-3}$ & $0.02 \times 10^{-3}$ & $0.85\times10^{-3}$ & $(1.17 \pm 0.04)\times10^{-3}$ & $0.43\times10^{-3}$ & - \\
		$D^0 \to \rho^- \pi^+$ & $1.00 \times 10^{-3}$ & $1.32\times10^{-3}$ & $1.52 \times 10^{-3}$ & $2.00\times10^{-3}$ & $(5.15 \pm 0.26)\times10^{-3}$ & $2.02\times10^{-3}$ & - \\
		$D^0 \to \eta \phi$ & $0.03 \times 10^{-4}$ & $1.06\times10^{-4}$ & $0.12 \times 10^{-4}$ & $3.83\times10^{-4}$ & $(1.80 \pm 0.5)\times10^{-4}$ & $3.87\times10^{-4}$ & - \\
		$D^0 \to K^- \rho^+$ & $0.45 \times 10^{-1}$ & $0.61\times10^{-1}$ & $0.79 \times 10^{-1}$ & $1.06\times10^{-1}$ & $(1.12 \pm 0.07)\times10^{-1}$ & $1.06\times10^{-1}$ & - \\
		$D^0 \to \eta \bar{K}^{*0}$ & $0.01 \times 10^{-2}$ & $0.43\times10^{-2}$ & $0.03 \times 10^{-2}$ & $0.98\times10^{-2}$ & $(1.41 \pm 0.12)\times10^{-2}$ & $0.99\times10^{-2}$ & - \\
		$D^0 \to \eta^{\prime} \bar{K}^{*0}$ & $0.06 \times 10^{-3}$ & $2.27\times10^{-3}$ & $0.001 \times 10^{-3}$ & $0.05\times10^{-3}$ & $<1\times10^{-3}$ & $0.07\times10^{-3}$ & - \\
		$D^0 \to \rho^0 \rho^0$ & $0.10 \times 10^{-3}$ & $3.20\times10^{-3}$ & $0.17 \times 10^{-3}$ & $5.49\times10^{-3}$ & $(1.85 \pm 0.13)\times10^{-3}$ & $0.69\times10^{-3}$ & - \\
		$D^0 \to \omega \phi$ & $0.31 \times 10^{-4}$ & $9.72\times10^{-4}$ & $0.29 \times 10^{-4}$ & $9.33\times10^{-4}$ & $(6.5 \pm 1.0)\times10^{-4}$ & $4.68\times10^{-4}$ & - \\
		\bottomrule
	\end{tabular}}
	\end{table}
	\newpage
	% Table generated by Excel2LaTeX from sheet 'Ds_n=3_included'
	\begin{table}[h]
		\tbl{Branching fractions for $D_s$ decays and their comparison with other references. \label{DsBfs}}
	{\begin{tabular}{lccccccc}
		\toprule
		\multicolumn{1}{c}{\multirow{3}[6]{*}{Channel}} & \multicolumn{4}{c}{This Work  } & \multirow{3}[6]{*}{PDG \cite{PDG2024}} & \multirow{3}[6]{*}{Ref.\citen{Galkin2023}} & \multirow{3}[6]{*}{Ref.\citen{PhysRevD.92.014032}} \\
		\cmidrule{2-5}      & \multicolumn{2}{c}{Hydrogenic} & \multicolumn{2}{c}{Gaussian} &       &       &  \\
		\cmidrule{2-5}      & $N=3$ & $N\to\infty$ & $N=3$ & $N\to\infty$ &       &       &  \\
		\midrule
		$D_s \to K^+ \bar{K}^0$ & $0.03 \times 10^{-2}$ & $1.13\times10^{-2}$ & $0.05 \times 10^{-2}$ & $1.57\times10^{-2}$ & $(2.95 \pm 0.14)\times10^{-2}$ & $1.57\times10^{-2}$ & - \\
		$D_s \to \eta \pi^+$ & $1.59 \times 10^{-2}$ & $2.12\times10^{-2}$ & $2.18 \times 10^{-2}$ & $2.92\times10^{-2}$ & $(1.67 \pm 0.09)\times10^{-2}$ & $2.92\times10^{-2}$ & $ (2.26 \pm 0.82)\times10^{-2} $ \\
		$D_s \to \eta' \pi^+$ & $1.24 \times 10^{-2}$ & $1.65\times10^{-2}$ & $1.96 \times 10^{-2}$ & $2.62\times10^{-2}$ & $(3.94 \pm 0.25)\times10^{-2}$ & $2.62\times10^{-2}$ & $ (2.64 \pm 0.78)\times10^{-2} $ \\
		$D_s \to K^+ \pi^0$ & $0.14 \times 10^{-4}$ & $4.56\times10^{-4}$ & $0.19 \times 10^{-4}$ & $6.22\times10^{-4}$ & $(7.4 \pm 0.50)\times10^{-4}$ & $3.16\times10^{-4}$ & $ (8.17 \pm 4.6)\times10^{-4} $ \\
		$D_s \to \eta K^+$ & $1.44 \times 10^{-3}$ & $4.50\times10^{-3}$ & $2.02 \times 10^{-3}$ & $6.32\times10^{-3}$ & $(1.73 \pm 0.08)\times10^{-3}$ & $3.97\times10^{-3}$ & $ (1.5 \pm 0.75)\times10^{-3} $ \\
		$D_s \to \eta^{\prime} K^+$ & $0.64 \times 10^{-3}$ & $0.47\times10^{-3}$ & $1.11 \times 10^{-3}$ & $0.83\times10^{-3}$ & $(2.64 \pm 0.24)\times10^{-3}$ & $0.30\times10^{-3}$ & $ (7.07 \pm 0.5)\times10^{-4}$ \\
		$D_s \to \eta \rho^+$ & $2.72 \times 10^{-2}$ & $3.63\times10^{-2}$ & $4.64 \times 10^{-2}$ & $6.21\times10^{-2}$ & $(8.9 \pm 0.8)\times10^{-2}$ & $6.20\times10^{-2}$ & - \\
		$D_s \to \eta^{\prime} \rho^+$ & $0.30 \times 10^{-2}$ & $0.40\times10^{-2}$ & $2.10 \times 10^{-2}$ & $2.81\times10^{-2}$ & $(5.8 \pm 1.5)\times10^{-2}$ & $2.79\times10^{-2}$ & - \\
		$D_s \to K^+ \omega$ & $0.28 \times 10^{-4}$ & $9.04\times10^{-4}$ & $0.47 \times 10^{-4}$ & $15.0\times10^{-4}$ & $(9.90 \pm 1.50)\times10^{-4}$ & $7.59\times10^{-4}$ & - \\
		$D_s \to K^+ \rho^0$ & $0.03 \times 10^{-3}$ & $1.09\times10^{-3}$ & $0.06 \times 10^{-3}$ & $1.79\times10^{-3}$ & $(2.17 \pm 0.25)\times10^{-3}$ & $0.91\times10^{-3}$ & - \\
		$D_s \to \phi \pi^+$ & $1.68 \times 10^{-2}$ & $2.25\times10^{-2}$ & $2.80 \times 10^{-2}$ & $3.74\times10^{-2}$ & $(4.5 \pm 0.4)\times10^{-2}$ & $3.73\times10^{-2}$ & - \\
		$D_s \to \phi \rho^+$ & $1.66 \times 10^{-2}$ & $2.22\times10^{-2}$ & $10.2 \times 10^{-2}$ & $13.6\times10^{-2}$ & $(5.59 \pm 0.34)\times10^{-2}$ & $13.3\times10^{-2}$ & - \\
		\bottomrule
	\end{tabular}}
\end{table}
	\newpage
%	 Table generated by Excel2LaTeX from sheet 'Sheet2'
	\begin{table}[h]
		\tbl{Branching fractions for $B$ decays and their comparison with other references.\label{BBfs}}
		{\begin{tabular}{lccccccc}
				\toprule
				\multicolumn{1}{c}{\multirow{3}[6]{*}{Channel}} & \multicolumn{4}{c}{This Work  } & \multirow{3}[6]{*}{PDG \cite{PDG2024}} & \multirow{3}[6]{*}{Ref.\citen{Galkin2023}} & \multirow{3}[6]{*}{Ref.\citen{PhysRevD.60.094014}} \\
				\cmidrule{2-5}      & \multicolumn{2}{c}{Hydrogenic} & \multicolumn{2}{c}{Gaussian} &       &       &  \\
				\cmidrule{2-5}      & $N=3$ & $N\to\infty$ & $N=3$ & $N\to\infty$ &       &       &  \\
				\midrule
				$B^+ \to \pi^+ \eta$ & $4.48 \times   10^{-6}$ & $3.36 \times 10^{-6}$ & $4.81 \times   10^{-6}$ & $3.60 \times 10^{-6}$ & $ (4.02 \pm 0.27)\times 10^{-6} $ & $3.79 \times 10^{-6}$ & - \\
				$B^+ \to \pi^+ \eta^{\prime}$ & $3.68 \times 10^{-6}$ & $3.00 \times 10^{-6}$ & $3.97 \times 10^{-6}$ & $3.23 \times 10^{-6}$ & $ (2.7 \pm 0.9)\times 10^{-6} $ & $3.32 \times 10^{-6}$ & - \\
				$B^+ \to \pi^+ \omega$ & $7.49 \times 10^{-6}$ & $3.58 \times 10^{-6}$ & $8.06 \times 10^{-6}$ & $3.85 \times 10^{-6}$ & $ (6.9 \pm 0.5)\times 10^{-6} $ & $1.92 \times 10^{-6}$ & - \\
				$B^+ \to \rho^+ \eta$ & $9.67 \times 10^{-6}$ & $8.93 \times 10^{-6}$ & $10.4 \times 10^{-6}$ & $9.62 \times 10^{-6}$ & $ (7.0 \pm 2.9)\times 10^{-6} $ & $9.31 \times 10^{-6}$ & - \\
				$B^+ \to \rho^+ \eta^{\prime}$ & $8.34 \times 10^{-6}$ & $7.99 \times 10^{-6}$ & $9.02 \times 10^{-6}$ & $8.64 \times 10^{-6}$ & $ (9.7 \pm 2.2)\times 10^{-6} $ & $8.43 \times 10^{-6}$ & - \\
				$B^+ \to \pi^+ K^0$ & $0.37 \times 10^{-5}$ & $0.10 \times 10^{-5}$ & $0.39 \times 10^{-5}$ & $0.11 \times 10^{-5}$ & $ (2.39 \pm 0.06)\times 10^{-5} $ & $0.52 \times 10^{-5}$ & $1.89\times 10^{-5}$ \\
				$B^+ \to \rho^+ K^0$ & $0.29 \times 10^{-6}$ & $0.70 \times 10^{-6}$ & $0.32 \times 10^{-6}$ & $0.75 \times 10^{-6}$ & $ (7.3^{+1.0}_{-1.2})\times 10^{-6} $ & $0.24 \times 10^{-6}$ & $0.24 \times 10^{-6}$ \\
				$B^+ \to \pi^+ \pi^0$ & $5.89 \times 10^{-6}$ & $3.59 \times 10^{-6}$ & $6.31 \times 10^{-6}$ & $3.84 \times 10^{-6}$ & $ (5.31 \pm 0.26)\times 10^{-6} $ & $1.89 \times 10^{-6}$ & - \\
				$B^+ \to \pi^+ \rho^0$ & $7.11 \times 10^{-6}$ & $2.84 \times 10^{-6}$ & $7.64 \times 10^{-6}$ & $3.05 \times 10^{-6}$ & $ (8.3 \pm 1.2)\times 10^{-6} $ & $1.55 \times 10^{-6}$ & - \\
				$B^+ \to \rho^+ \pi^0$ & $1.34 \times 10^{-5}$ & $1.05 \times 10^{-5}$ & $1.44 \times 10^{-5}$ & $1.13 \times 10^{-5}$ & $ (1.06^{+0.12}_{-0.13})\times 10^{-5} $ & $0.55 \times 10^{-5}$ & - \\
				$B^+ \to \pi^+ \phi$ & $0.22 \times 10^{-8}$ & $5.14 \times 10^{-8}$ & $0.24 \times 10^{-8}$ & $5.54 \times 10^{-8}$ & $ (3.2 \pm 1.5)\times 10^{-8} $ & $5.52 \times 10^{-8}$ & - \\
				$B^+ \to \rho^+ \rho^0$ & $1.80 \times 10^{-5}$ & $1.10 \times 10^{-5}$ & $1.90 \times 10^{-5}$ & $1.16 \times 10^{-5}$ & $ (2.40 \pm 0.19) \times 10^{-5} $ & $0.67 \times 10^{-5}$ & - \\
				$B^+ \to \rho^+ \omega$ & $1.82 \times 10^{-5}$ & $1.19 \times 10^{-5}$ & $1.95 \times 10^{-5}$ & $1.28 \times 10^{-5}$ & $ (1.59 \pm 0.21)\times 10^{-5} $ & $0.71 \times 10^{-5}$ & - \\
				$B^0 \to D^{-} \pi^+$ & $4.07 \times 10^{-3}$ & $5.00 \times 10^{-3}$ & $4.13 \times 10^{-3}$ & $5.07 \times 10^{-3}$ & $ (2.51 \pm 0.08)\times 10^{-3} $ & $4.80 \times 10^{-3}$ & - \\
				$B^0 \to D^{-} K^{+}$ & $3.13 \times 10^{-4}$ & $3.79 \times 10^{-4}$ & $3.17 \times 10^{-4}$ & $3.85 \times 10^{-4}$ & $ (2.05 \pm 0.08)\times 10^{-4} $ & $3.69 \times 10^{-4}$ & - \\
				$B^0 \to \pi^- K^{+}$ & $0.31 \times 10^{-5}$ & $0.07 \times 10^{-5}$ & $0.33 \times 10^{-5}$ & $0.07 \times 10^{-5}$ & $ (2.00 \pm 0.04)\times 10^{-5} $ & $0.40 \times 10^{-5}$ & $1.56 \times 10^{-5}$ \\
				$B^0 \to \pi^- \pi^+$ & $6.12 \times 10^{-6}$ & $7.50 \times 10^{-6}$ & $6.56 \times 10^{-6}$ & $8.04 \times 10^{-6}$ & $ (5.37 \pm 0.20)\times 10^{-6} $ & $5.65 \times 10^{-6}$ & - \\
				$B^0 \to \pi^0 \pi^0$ & $0.49 \times 10^{-6}$ & $0.93 \times 10^{-6}$ & $0.52 \times 10^{-6}$ & $1.00 \times 10^{-6}$ & $ (1.55 \pm 0.17)\times 10^{-6} $ & $0.15 \times 10^{-6}$ & $0.90\times 10^{-6}$ \\
				$B^0 \to \pi^0 \eta$ & $1.72 \times 10^{-7}$ & $4.89 \times 10^{-7}$ & $1.84 \times 10^{-7}$ & $5.24 \times 10^{-7}$ & $ (4.1 \pm 1.7)\times 10^{-7} $ & $1.33 \times 10^{-7}$ & - \\
				$B^0 \to \pi^0 \eta^{\prime}$ & $0.13 \times 10^{-6}$ & $0.37 \times 10^{-6}$ & $0.14 \times 10^{-6}$ & $0.40 \times 10^{-6}$ & $ (1.2 \pm 0.6)\times 10^{-6} $ & $0.07 \times 10^{-6}$ & $0.07\times 10^{-6}$ \\
				$B^0 \to \eta \eta$ & $0.24 \times 10^{-6}$ & $0.30 \times 10^{-6}$ & $0.26 \times 10^{-6}$ & $0.32 \times 10^{-6}$ & $ <1.0\times10^{-6} $ & $0.51 \times 10^{-6}$ & - \\
				$B^0 \to \eta^{\prime} \eta^{\prime}$ & $0.13 \times 10^{-6}$ & $0.14 \times 10^{-6}$ & $0.14 \times 10^{-6}$ & $0.16 \times 10^{-6}$ & $ <1.7\times10^{-6} $ & $0.21 \times 10^{-6}$ & - \\
				$B^0 \to D^{*-} \pi^+$ & $4.40 \times 10^{-3}$ & $5.41 \times 10^{-3}$ & $5.24 \times 10^{-3}$ & $6.43 \times 10^{-3}$ & $ (2.66 \pm 0.07)\times 10^{-3} $ & $6.13 \times 10^{-3}$ & - \\
				$B^0 \to D^{-} \rho^+$ & $9.80 \times 10^{-3}$ & $12.0 \times 10^{-3}$ & $9.94 \times 10^{-3}$ & $12.2 \times 10^{-3}$ & $ (7.6 \pm 1.2)\times 10^{-3} $ & $11.6 \times 10^{-3}$ & - \\
				$B^0 \to \rho^0 \pi^0$ & $0.53 \times 10^{-6}$ & $1.57 \times 10^{-6}$ & $0.57 \times 10^{-6}$ & $1.68 \times 10^{-6}$ & $ (2.0 \pm 0.5)\times 10^{-6} $ & $0.40 \times 10^{-6}$ & $0.03\times 10^{-6}$ \\
				$B^0 \to \omega \pi^0$ & $5.74 \times 10^{-7}$ & $14.5 \times 10^{-7}$ & $6.17 \times 10^{-7}$ & $15.6 \times 10^{-7}$ & $ <5.0\times 10^{-7} $ & $0.14 \times 10^{-7}$ & - \\
				$B^0 \to \rho^- K^{+}$ & $0.76 \times 10^{-6}$ & $0.56 \times 10^{-6}$ & $0.82 \times 10^{-6}$ & $0.60 \times 10^{-6}$ & $ (7.0 \pm 0.9)\times 10^{-6} $ & $1.08 \times 10^{-6}$ & $1.16 \times 10^{-6}$ \\
				$B^0 \to \pi^- K^{*+}$ & $1.06 \times 10^{-6}$ & $1.18 \times 10^{-6}$ & $1.14 \times 10^{-6}$ & $1.27 \times 10^{-6}$ & $ (7.5 \pm 0.4)\times 10^{-6} $ & $1.30 \times 10^{-6}$ & $6.84 \times 10^{-6}$ \\
				$B^0 \to \rho^- \pi^+$ & $0.45 \times 10^{-5}$ & $0.56 \times 10^{-5}$ & $0.48 \times 10^{-5}$ & $0.60 \times 10^{-5}$ & $ (2.30 \pm 0.23)\times 10^{-5} $ & $0.59 \times 10^{-5}$ & $0.81 \times 10^{-5}$ \\
				$B^0 \to D^{-} K^{*+}$ & $5.87 \times 10^{-4}$ & $7.12 \times 10^{-4}$ & $5.96 \times 10^{-4}$ & $7.22 \times 10^{-4}$ & $ (4.5 \pm 0.7)\times 10^{-4} $ & $6.93 \times 10^{-4}$ & - \\
				$B^0 \to D^{*-} K^{+}$ & $3.31 \times 10^{-4}$ & $4.01 \times 10^{-4}$ & $3.96 \times 10^{-4}$ & $4.80 \times 10^{-4}$ & $ (2.16 \pm 0.08)\times 10^{-4} $ & $4.64 \times 10^{-4}$ & - \\
				$B^0 \to \eta^{\prime} \eta$ & $0.14 \times 10^{-6}$ & $0.21 \times 10^{-6}$ & $0.15 \times 10^{-6}$ & $0.22 \times 10^{-6}$ & $ <1.2\times 10^{-6} $ & $0.33 \times 10^{-6}$ & - \\
				$B^0 \to \rho^0 \eta$ & $0.33 \times 10^{-6}$ & $0.89 \times 10^{-6}$ & $0.36 \times 10^{-6}$ & $0.95 \times 10^{-6}$ & $ <1.5\times 10^{-6} $ & $0.07 \times 10^{-6}$ & - \\
				$B^0 \to \rho^0 \eta^{\prime}$ & $0.25 \times 10^{-6}$ & $0.72 \times 10^{-6}$ & $0.27 \times 10^{-6}$ & $0.78 \times 10^{-6}$ & $ <1.3\times 10^{-6} $ & $0.07 \times 10^{-6}$ & - \\
				$B^0 \to \omega \eta$ & $4.45 \times 10^{-7}$ & $8.16 \times 10^{-7}$ & $4.80 \times 10^{-7}$ & $8.79 \times 10^{-7}$ & $ (9.4^{+4.0}_{-3.1})\times 10^{-7} $ & $4.32 \times 10^{-7}$ & - \\
				$B^0 \to \omega \eta^{\prime}$ & $0.29 \times 10^{-6}$ & $0.65 \times 10^{-6}$ & $0.31 \times 10^{-6}$ & $0.70 \times 10^{-6}$ & $ (1.0^{+0.5}_{-0.4})\times 10^{-6} $ & $0.35 \times 10^{-6}$ & - \\
				$B^0 \to \pi^- \rho^+$ & $1.04 \times 10^{-5}$ & $1.30 \times 10^{-5}$ & $1.12 \times 10^{-5}$ & $1.40 \times 10^{-5}$ & $ (2.30 \pm 0.23)\times 10^{-5} $ & $1.37 \times 10^{-5}$ & - \\
				$B^0 \to \rho^- \rho^+$ & $1.29 \times 10^{-5}$ & $1.62 \times 10^{-5}$ & $1.37 \times 10^{-5}$ & $1.71 \times 10^{-5}$ & $ (2.77 \pm 0.19)\times 10^{-5} $ & $1.67 \times 10^{-5}$ & - \\
				$B^0 \to \rho^0 \rho^0$ & $10.0 \times 10^{-7}$ & $28.4 \times 10^{-7}$ & $10.6 \times 10^{-7}$ & $30.0 \times 10^{-7}$ & $ (9.6 \pm 1.5)\times 10^{-7} $ & $0.36 \times 10^{-7}$ & $0.50 \times 10^{-7}$ \\
				$B^0 \to \omega \omega$ & $0.98 \times 10^{-6}$ & $2.38 \times 10^{-6}$ & $1.05 \times 10^{-6}$ & $2.55 \times 10^{-6}$ & $ (1.2 \pm 0.4)\times 10^{-6} $ & $0.01 \times 10^{-6}$ & $0.07 \times 10^{-6}$ \\
				$B^0 \to \omega \rho^0$ & $0.90 \times 10^{-6}$ & $2.61 \times 10^{-6}$ & $0.95 \times 10^{-6}$ & $2.77 \times 10^{-6}$ & $ <1.6\times 10^{-6} $ & $0.03 \times 10^{-6}$ & - \\
				$B^0 \to D^{*-} \rho^+$ & $11.8 \times 10^{-3}$ & $14.5 \times 10^{-3}$ & $14.2 \times 10^{-3}$ & $17.5 \times 10^{-3}$ & $ (6.8 \pm 0.9)\times 10^{-3} $ & $16.7 \times 10^{-3}$ & - \\
				$B^0 \to D^{*-} K^{*+}$ & $7.33 \times 10^{-4}$ & $8.89 \times 10^{-4}$ & $8.85 \times 10^{-4}$ & $10.7 \times 10^{-4}$ & $ (3.3 \pm 0.6)\times 10^{-4} $ & $10.4 \times 10^{-4}$ & - \\
				\bottomrule
			\end{tabular}}
	\end{table}
	\newpage
%	 Table generated by Excel2LaTeX from sheet 'Sheet3'
	\begin{table}[h]
		\tbl{Branching fractions for $B_s$ decays and their comparison with other references \label{BsBfs}}
		{% Table generated by Excel2LaTeX from sheet 'Bs_n=3_included'
			\begin{tabular}{lccccccc}
				\toprule
				\multicolumn{1}{c}{\multirow{3}[6]{*}{Channel}} & \multicolumn{4}{c}{This Work  } & \multirow{3}[6]{*}{PDG \cite{PDG2024}} & \multirow{3}[6]{*}{Ref.\citen{Galkin2023}} & \multirow{3}[6]{*}{Ref.\citen{PhysRevD.59.074003}} \\
				\cmidrule{2-5}      & \multicolumn{2}{c}{Hydrogenic} & \multicolumn{2}{c}{Gaussian} &       &       &  \\
				\cmidrule{2-5}      & $N=3$ & $N\to \infty$ & $N=3$ & $N\to \infty$ &       &       &  \\
				\midrule
				$B_s \to D^{-}_s \pi^+$ & $3.78   \times 10^{-3}$ & $4.65 \times 10^{-3}$ & $3.82 \times   10^{-3}$ & $4.69 \times 10^{-3}$ & $ (2.98 \pm 0.14)\times 10^{-3} $ & $4.43 \times 10^{-3}$ & - \\
				$B_s \to D^{-}_s K^{+}$ & $2.92 \times 10^{-4}$ & $3.54 \times 10^{-4}$ & $2.94 \times 10^{-4}$ & $3.57 \times 10^{-4}$ & $ (2.25 \pm 0.12)\times 10^{-4} $ & $3.41 \times 10^{-4}$ & - \\
				$B_s \to \pi^+ K^{-}$ & $7.62 \times 10^{-6}$ & $9.34 \times 10^{-6}$ & $8.16 \times 10^{-6}$ & $10.0 \times 10^{-6}$ & $ (5.9 \pm 0.7)\times 10^{-6} $ & $9.91 \times 10^{-6}$ & - \\
				$B_s \to K^{+} K^{-}$ & $0.55 \times 10^{-5}$ & $0.13 \times 10^{-5}$ & $0.59 \times 10^{-5}$ & $0.14 \times 10^{-5}$ & $ (2.72 \pm 0.23)\times 10^{-5} $ & $0.69 \times 10^{-5}$ & - \\
				$B_s \to D^{-}_s \rho^+$ & $9.14 \times 10^{-3}$ & $11.2 \times 10^{-3}$ & $9.21 \times 10^{-3}$ & $11.3 \times 10^{-3}$ & $ (6.8 \pm 1.4)\times 10^{-3} $ & $10.7 \times 10^{-3}$ & - \\
				$B_s \to D^{*-}_s \pi^+$ & $2.66 \times 10^{-3}$ & $3.26 \times 10^{-3}$ & $3.14 \times 10^{-3}$ & $3.86 \times 10^{-3}$ & $ (1.9^{+0.5}_{-0.4})\times 10^{-3} $ & $3.66 \times 10^{-3}$ & - \\
				$B_s \to D^{*-}_s K^{+}$ & $1.99 \times 10^{-4}$ & $2.42 \times 10^{-4}$ & $2.37 \times 10^{-4}$ & $2.88 \times 10^{-4}$ & $ (1.32^{+0.40}_{-0.32})\times 10^{-4} $ & $2.76 \times 10^{-4}$ & - \\
				$B_s \to D^{-}_s K^{*+}$ & $5.48 \times 10^{-4}$ & $6.64 \times 10^{-4}$ & $5.52 \times 10^{-4}$ & $6.69 \times 10^{-4}$ & -     & $6.41 \times 10^{-4}$ & - \\
				$B_s \to K^{*-} \pi^+$ & $7.27 \times 10^{-6}$ & $9.09 \times 10^{-6}$ & $7.81 \times 10^{-6}$ & $9.76 \times 10^{-6}$ & $ (2.9 \pm 1.1)\times10^{-6} $ & $9.54 \times 10^{-6}$ & - \\
				$B_s \to K^{*-} K^{+}$ & $0.12 \times 10^{-5}$ & $0.08 \times 10^{-5}$ & $0.13 \times 10^{-5}$ & $0.10 \times 10^{-5}$ & $ (1.9 \pm 0.5)\times 10^{-5} $ & $0.17\times 10^{-5}$ & $7.5\times10^{-7}$ \\
				$B_s \to K^{-} K^{*+}$ & $1.79 \times 10^{-6}$ & $1.99 \times 10^{-6}$ & $1.93 \times 10^{-6}$ & $2.15 \times 10^{-6}$ & $ (1.9 \pm 0.5)\times 10^{-6} $ & $2.20 \times 10^{-6}$ & $3.77\times10^{-6}$ \\
				$B_s \to D^{*-}_s K^{*+}$ & $5.10 \times 10^{-4}$ & $6.18 \times 10^{-4}$ & $5.87 \times 10^{-4}$ & $7.11 \times 10^{-4}$ & - & $6.81 \times 10^{-4}$ & - \\
				$B_s \to D^{*-}_s \rho^+$ & $8.03 \times 10^{-3}$ & $9.87 \times 10^{-3}$ & $9.25 \times 10^{-3}$ & $11.4 \times 10^{-3}$ & $ (9.5 \pm 2.0)\times 10^{-3} $ & $10.7 \times 10^{-3}$ & - \\
				\bottomrule
			\end{tabular}}
	\end{table}
	
	\noindent We compare the branching fractions obtained using Gaussian predicted meson masses with PDG 2024 averages and some other references.\\
	In case of charm decays we find that the number of color tends to $\infty$ limit improves the theoretical prediction as pointed out in reference \citen{Galkin2023}.\\
	We observe several channels where the hydrogenic predicted mass values yield more accurate branching fractions than the Gaussian ones, including $D^+ \to \pi^0 \pi^+$, $D^+ \to \pi^0 K^+$, $D^+ \to \eta K^+$, $D^0 \to \pi^0 \pi^0$, $D_s^+ \to \eta \pi^+$, and $D_s^+ \to \phi \rho^+$. While the Gaussian model reproduces experimental charm meson masses with high precision, the hydrogenic model noticeably underestimates them (e.g., yielding $M_D \approx$ 1.702 GeV compared to the experimental 1.869 GeV). To understand this apparent contradiction, we must look at the complete kinematic dependence of the branching fraction. The general formula of $\mathcal{B}$ includes a factor of $|\vec{p}|/m_M^2$. However, the decay amplitude $\mathcal{A}$ for $M \to PP$ transitions (Eq. (\ref{PP})) is proportional to $m_M^2$. Because the branching fraction scales with $|\mathcal{A}|^2$, the overall dependence becomes $\mathcal{B} \propto |\vec{p}| m_M^2$. Additionally, when the masses of the daughter particles are small or comparable with each other, the momentum $|\vec{p}|$ derived from Eq. (\ref{kallen}) becomes roughly constant. Therefore, the $m_M$ contribution dominates the final branching fraction. We can say that the branching fraction is proportional to mass value of the parent particle. In a sense that an increment in the mass value implies an increment in $\mathcal{B}$. This trend is evident in the PV, VP, and VV decay modes. Consequently, replacing the physical mass with the artificially lower hydrogenic mass severely restricts the available phase space and suppresses the calculated branching fraction. The necessity of this mathematical suppression highlights a fundamental limitation: naive factorization systematically overestimates the theoretical amplitudes for these specific charm channels. The physical origin of this overestimation is grounded in the space-time evolution of the decay. As noted by Bjorken\cite{BJORKEN1989325}, the factorization hypothesis is widely used for bottom decays but ``less justifiably" applied to charm decays. Factorization relies on the decay products possessing extreme relativistic recoil; the resulting relativistic time-dilation ensures that the daughter mesons have a long formation time, allowing them to form several fermis away from the parent's color field and thus avoid final-state interactions. Charm decays lack this extreme recoil energy. The daughter mesons hadronize while still entangled in the spectator system's color fields, making them highly susceptible to non-factorizable soft-gluon exchanges and final-state interactions. In color-suppressed and interference-dominated channels, naive factorization fails to account for these complex dynamics, frequently leading to artificially inflated decay amplitudes.Therefore, the success of the hydrogenic mass in these modes is not coincidental; the restricted phase space acts as a necessary kinematic regulator. It artificially scales down the overpredicted factorization amplitudes to cleanly offset the missing physical suppression, bringing the calculated branching fractions into alignment with experimental data. Conversely, for tree-dominated transitions (such as $D^+ \to \eta \omega$ or $D_s^+ \to K^+ \rho^0$), the naive factorization amplitudes are more accurate. In these well-behaved channels, the suppression from the hydrogenic mass would excessively diminish the rate, meaning the physically accurate Gaussian mass is required to provide the correct kinematic normalization.\\
	Interestingly, the hydrogenic-inspired mass does not always predict lower branching fractions in the charm sector compared to the Gaussian model; higher values can occur when the parent mass in the denominator dominates the $|\vec{p}|$ calculation, a kinematic effect particularly evident in near-threshold decays like $D^0 \to \omega \phi$ ($m_\omega + m_\phi = 0.782 + 1.020 = 1.802 \text{ GeV}$) and $D^0 \to \eta^{\prime} \bar{K}^{*0}$ ($m_{\eta'} + m_{\bar{K}^{*0}} = 0.959 + 0.891 = 1.850 \text{ GeV}$), where the combined daughter masses closely approach the parent mass.\\
	In the case of bottom meson decays, we find that the branching fractions calculated with $N=3$ are in good agreement with the PDG 2024 values. The $N\to\infty$ limit does not yield an improvement in the non-factorizable contributions, a behavior that is expected given that the factorization approach is formulated for heavy mesons \cite{BJORKEN1989325}.For bottom mesons, the masses predicted using Gaussian and hydrogen-like wavefunctions differ by only $\approx 2\%$; consequently, significant deviations are $5-20\%$. Nevertheless, because the factorization framework is explicitly designed for heavy mesons, it is crucial to investigate how these minor theoretical mass deviations impact the decay rates. A primary objective of this work is to utilize theoretical mass values to study the sensitivity of the branching fractions to such variations. Although a $5-20\%$ variation in theoretical branching fractions has historically been considered acceptable within naive factorization, in the present context it reveals a pronounced non-linear sensitivity of the decay rates to the parent mass input. We observe that a relative mass shift of order $2–10\%$ between the Gaussian and hydrogenic predictions can induce up to a $100\%$ change in the corresponding branching fractions.\\
	Using the factorization framework allows us to compare our branching fractions directly with established literature. We evaluate our results against Refs. \citen{PhysRevD.92.014032, PhysRevD.60.094014, PhysRevD.59.074003} to assess the limitations of the pure factorization approach. In particular, comparing our work with Ref. \citen{PhysRevD.60.094014} shows the effect of using more advanced form factors. Ultimately, this analysis confirms that deviations from experimental data in specific modes occur across these models, mainly because they do not include strong final-state interactions. For detailed analysis refer to Ref. \citen{Galkin2023}.
	This amplification arises from the nonlinear dependence of the phase-space factor and interference-sensitive amplitudes on the parent meson mass. In the era of precision measurements at experiments such as LHCb and Belle II, such sensitivity implies that the choice of mass input — and consequently the underlying meson wavefunction — is not merely technical but can significantly impact phenomenological predictions. Since the Gaussian approach reproduces the experimentally measured masses more accurately, it provides a more stable baseline for branching-fraction calculations. It is one of the simplest theoretical methods capable of yielding accurate mass predictions for mesons, including their excited states, as demonstrated in previous studies by our group \cite{devlani_mass_2013, devlani_spectroscopy_2011, devlani_spectroscopy_2012}. For theoretical treatments of the $B_c$ meson, one may refer to the Gaussian wave function approach\cite{Bc1} or the pNRQCD framework\cite{Bc2}. Given the simplicity of the factorization approach combined with these reliable mass predictions, this formalism can be readily extended to predict the decay properties of exotic particles—such as the excited states of $B_c$ mesons and $T_{bb}$ tetraquarks—where experimental masses are currently unavailable. For comprehensive details on extending the Gaussian approach to other hadronic systems, readers are directed to Ref.\citen{GEM}.
        
    \section{Conclusion}
        The primary objective of this study was to investigate the sensitivity of the naive factorization framework to variations in theoretical parent-meson mass inputs. Rather than treating the mass as a fixed empirical parameter, we utilized purely theoretical masses derived from a Coulomb plus linear potential model evaluated with both hydrogenic and Gaussian wavefunctions. Previous work has well established that the Gaussian-inspired models reproduce experimental heavy-light meson masses with high precision, providing a reliable theoretical baseline.\\
        By comparing the branching fractions calculated from these distinct mass inputs, we observed a profound, non-linear sensitivity within the factorization approach. A mass variation of order $2-10\%$ induced shifts of up to $100\%$ in the corresponding branching fractions. In the bottom sector ($B$ and $B_s$ mesons), predictions utilizing the precise Gaussian-like masses and $N = 3$ demonstrated good agreement with current experimental averages. The $N \to \infty$ limit offered no improvement, aligning with the expectation that naive factorization is fundamentally well-suited for the bottom sector without requiring artificial large-$N$ compensation.\\
        Conversely, our channel-by-channel analysis of the charm sector ($D$ and $D_s$ mesons) revealed fundamental limitations of the factorization approximation. In charm decays, the daughter mesons lack extreme relativistic recoil, making them highly susceptible to final-state interactions and non-factorizable soft-gluon exchanges. While taking the $N \to \infty$ limit partially compensated for these effects, we found that the artificially lower hydrogenic mass predictions yielded accurate branching fractions for several color-suppressed channels. This phenomenon demonstrates that the restricted phase space enforced by the underestimated hydrogenic mass acts as a necessary kinematic regulator, offsetting the inflated amplitudes inherent to charm factorization.\\
        Ultimately, this sensitivity analysis establishes a powerful benchmark. By combining the mathematically simple factorization approach with the reasonably accurate Gaussian mass predictions, we have validated a purely theoretical formalism capable of producing reliable decay rates without requiring prior experimental mass inputs. This framework can now be confidently extended to predict the decay properties of unobserved or exotic hadrons, such as the excited states of $B_c$ mesons and $T_{bb}$ tetraquarks, providing critical guidance for future measurements at facilities like LHCb and Belle II.
	\bibliographystyle{ws-ijmpa}
	\bibliography{references}
	
\end{document}